\label{key}

\documentclass{jfm}

\usepackage{graphicx}
\usepackage{epstopdf,epsfig}
\usepackage{newtxtext}
\usepackage{newtxmath}
\usepackage{natbib}

\newcommand{\RomanNumeralCaps}[1]
\linenumbers

\usepackage{bm}%
\usepackage[colorlinks,
linkcolor=blue,
anchorcolor=blue,
citecolor=blue
]{hyperref}
\usepackage{subfigure}
\usepackage{multirow}
\usepackage{natbib}
\usepackage{float}
\usepackage{makecell}
\usepackage{amsmath}

\shorttitle{Near-wall model for compressible turbulent boundary layers}
\shortauthor{K. P. Griffin, L. Fu and P. Moin}

\title{Near-wall model for compressible turbulent boundary layers based on an inverse velocity transformation}

\author{Kevin P. Griffin\aff{1,2}$^,$\corresp{\email{kevinpg@stanford.edu}},
Lin Fu\aff{3}$^,$\corresp{\email{linfu@ust.hk}}, Parviz  Moin\aff{1}
 }

\affiliation{\aff{1}Center for Turbulence Research, Stanford University,
Stanford, CA 94305
\aff{2}National Renewable Energy Laboratory, Golden, CO 80401
\aff{3}Department of Mechanical and Aerospace Engineering and Department of Mathematics, The Hong Kong University of Science and Technology, Clear Water Bay, Kowloon, Hong Kong
}

\usepackage{todonotes}
\newcommand{\tder}[2]{\frac{\mathrm{d}#1}{\mathrm{d}#2}} 
\newcommand{\tderi}[2]{\mathrm{d}#1 / \mathrm{d}#2} 

\begin{document}

\maketitle

\begin{abstract} 
In this work, a near-wall model, which couples the inverse of a recently developed compressible velocity transformation [Griffin, Fu, \& Moin, {\it PNAS}, 118:34, 2021] and an algebraic temperature-velocity relation, is developed for high-speed turbulent boundary layers.
As input, the model requires the mean flow state at one wall-normal height in the inner layer of the boundary layer and at the boundary-layer edge. 
As output, the model can predict mean temperature and velocity profiles across the entire inner layer, as well as the wall shear stress and heat flux. 
The model is tested in an {\it a priori} sense using a wide database of direct numerical simulation high-Mach-number turbulent channel flows, pipe flows, and boundary layers (48 cases with edge Mach numbers in the range of 0.77--11 and semi-local friction Reynolds numbers in the range of 170--5700).
The present model is significantly more accurate than the classical ordinary differential equation (ODE) model for all cases tested.
The model is deployed as a wall model for large-eddy simulations in channel flows with bulk Mach numbers in the range of 0.7--4 and friction Reynolds numbers in the range of 320--1800. When compared to the classical framework, in the {\it a posteriori} sense, the present method greatly improves the predicted heat flux, wall stress, and temperature and velocity profiles, especially in cases with strong heat transfer. In addition, the present model solves one ODE instead of two and has a similar computational cost and implementation complexity as the commonly used ODE model.
\end{abstract}



\section{Introduction}
The largest driver of computational cost in numerical simulations of wall-bounded turbulence is typically the numerical resolution in the near-wall region. In scale-resolving simulations, e.g., wall-resolved (WR) large-eddy simulation (LES), high spatial and temporal resolutions are required to accurately simulate the small-scale eddies near walls. Wall models, or approximate boundary conditions, can be employed to reduce the near-wall resolution requirements.
The computational cost (the number of grid points multiplied by the number of time steps) for the simulation of a turbulent boundary layer scales with the Reynolds number as $Re^{2.7}$ for WRLES and $Re^{1.1}$ for wall-modeled (WM) LES \citep{Yang2021}. Thus, wall models lead to substantial cost savings for high-Reynolds-number applications.
In simulations of the Reynolds-averaged Navier-Stokes (RANS) equations, high spatial resolution is also required to resolve the steep near-wall gradients in the mean flow. 
Therefore, wall models \textemdash typically referred to as wall functions in the RANS context \textemdash can also greatly accelerate numerical simulations.

The present work focuses on the paradigm of wall-stress modeling \citep{Larsson2016,Bose2018} for LES. These models were derived from RANS analysis of boundary layers and typically invoke a zero-equation RANS model such as the Prandtl mixing length argument \citep{Prandtl1925}, which models the turbulence length scale as a linear function of the wall-normal distance. An empirical damping function is introduced following \cite{VanDriest1956} to ensure the correct near-wall scaling of the mixing length. RANS models have naturally been widely used as boundary conditions for under-resolved RANS simulations (e.g., \cite{Abrahamson1988,Lien1998,Goncalves2001,Parente2011a}). In this context, such a model is typically referred to as a wall function.
\cite{Cabot1995, Cabot2000} showed that the mixing length RANS model is suitable for use as a boundary condition for the LES equations, i.e., for deployment as a wall-stress model. Specifically, they invoke the
 one-dimensional simplification of the RANS streamwise momentum equation. That is,
\begin{equation} \label{eq:u_ode}
\tder{}{y}\left((\overline{\mu} + \overline{\mu}_t) \tder{\widetilde{U}}{y} \right) = 0,
\end{equation}
where $\overline{\mu}$, $\overline{\mu}_t$, and $\widetilde{U}$ are the molecular dynamic viscosity, eddy viscosity, and velocity profiles, respectively, and $y$ is the wall-normal coordinate. $\overline{(\cdot)}$ denotes the Reynolds average and $\widetilde{(\cdot)}$ denotes the Favre (density-weighted) average. Throughout this work the Favre- (density-weighted-) averaged RANS and LES equations are employed.
The eddy viscosity is further modeled as
\begin{equation} \label{eq:mu_t}
\overline{\mu}_t = \kappa y \overline{\rho} \sqrt{\tau_w/\overline{\rho}} \left( 1 - \exp (y^+/A^+) \right)^2,
\end{equation}
where $\overline{\rho}(y)$ is the density profile. The subscript $(\cdot)_w$ denotes quantities evaluated at the wall. $\tau_w = \overline{\mu}_w (d\widetilde{U}/dy)_w$ is the wall shear stress. The superscript $(\cdot)^+$ denotes non-dimensionalization by the friction velocity $u_\tau = \sqrt{\tau_w/\overline{\rho}_w}$, $\overline{\rho}_w$, and the kinematic wall viscosity $\overline{\nu}_w=\overline{\mu}_w/\overline{\rho}_w$. 

The von K{\'a}rm{\'a}n constant $\kappa = 0.41$ and the eddy-viscosity damping coefficient $A^+ = 17$ are adopted following \cite{Cabot2000}.

For an incompressible flow, the density and molecular dynamic viscosity are known constants. In the context of WMLES, the ODE in Eq.~(\ref{eq:u_ode}) is solved with two boundary conditions: 1) the no-slip wall condition and 2) a velocity sample, which is taken from the LES at a wall-normal distance referred to as the matching location. Note that the solution procedure is iterative because the eddy viscosity depends on the wall stress (Eq.~(\ref{eq:mu_t})). The computed wall stress $\tau_w$ is then applied as a momentum-flux boundary condition for the outer LES solver, which completes the two-way coupling of the wall model (inner) solution and the PDE (outer) simulation.

For compressible flow, the RANS equation for temperature can similarly be simplified to the one-dimensional form \citep{Larsson2016,Bose2018}, which results in a second, coupled ODE for the temperature profile, i.e.,
\begin{equation} \label{eq:T_ode}
	\tder{}{y}\left((\overline{\mu} + \overline{\mu}_t) \widetilde{U}\tder{\widetilde{U}}{y} + C_p (\frac{\overline{\mu}}{\Pr}+\frac{\overline{\mu}_t}{\Pr_t})\tder{\widetilde{T}}{y}\right) = 0,
\end{equation}
where $\widetilde{T}$ is the temperature profile. $C_p$ is the specific heat capacity at constant pressure, $\Pr$ is the Prandtl number, and $\Pr_t$ is the turbulent Prandtl number, which is assumed to be 0.9 \citep{Larsson2016}.
The dependence of molecular dynamic viscosity on temperature can be assumed to follow a power law or Sutherland's law.
The ideal gas equation of state closes the system
and the thin-boundary-layer assumption implies that the pressure is constant across the inner layer.

In WMLES, the temperature ODE in Eq.~(\ref{eq:T_ode}) is solved with two additional boundary conditions: 1) the wall temperature and 2) the temperature at the matching location. Note that the solution procedure is also iterative in that the temperature depends on the velocity solution. The velocity also depends on the temperature through the density and viscosity. Solving two coupled boundary-value problems iteratively introduces a higher degree of non-linearity compared to the incompressible case and can prove difficult to converge in flows with strong temperature gradients (strong heat transfer), e.g., as was reported in \cite{Fu2021}. In addition to the numerical difficulties, the accuracy of this wall model degrades substantially in flows with strong heat transfer (as will be demonstrated herein).

Improved results for high-speed wall-bounded turbulent flows over cold walls have been obtained by using the semi-local scaling in the damping function \citep{Yang2018a,Fu2022}, however, \cite{Iyer2019} reports that in adiabatic walls, the classical scaling (consistent with the van Driest transformation) is more accurate. This motivates using a recently developed compressible velocity transformation that is accurate for both diabatic and adiabatic turbulent boundary layers \citep{Griffin2021a}.

In this work, a wall model for high-speed wall-bounded turbulent flows is developed in section \ref{sec:model}. The model is evaluated via {\it a priori} testing in section \ref{sec:a_priori} and via {\it a posteriori} validation in section \ref{sec:a_posteriori}. Conclusions are drawn in section \ref{sec:conclusion}.

\section{Model development} \label{sec:model}
There are two principal differences between the present model and the classical ODE-based wall model (Eqs.~(\ref{eq:u_ode}--\ref{eq:T_ode})): (1) rather than solving an ODE for the compressible velocity profile directly, the incompressible ODE (with constant density and viscosity) is solved, and an inverse compressibility transformation \citep{Griffin2021a} is employed; (2) rather than employing a RANS equation for temperature and assuming a constant $Pr_t$, an algebraic temperature-velocity relation is adopted, thus obviating the need to solve a second ODE.

\subsection{Inverse compressible velocity transformation}
A compressible velocity transformation seeks to map the local mean strain rate of the variable-property compressible flow, $d\widetilde{U}/dy$, to the non-dimensional mean strain rate of a constant-property incompressible flow at an equivalent Reynolds number. Upon integration, the transformation maps the compressible velocity profile to an incompressible velocity profile. In this way, a successful transformation can collapse profiles with different Mach numbers and thermal boundary conditions to a single incompressible law of the wall. Coupled with the incompressible profile implied by Eq.~(\ref{eq:u_ode}), an inverse velocity transformation can recover the compressible velocity profile. 

The total-stress-based compressible velocity transformation of \cite{Griffin2021a} is used in this work since it is shown to be accurate in a wide range of flows, including boundary layers with strong heat transfer. This transformation uses the viscous scaling arguments of \cite{Trettel2016} and \cite{Patel2016} in the near-wall viscous region and uses a modified version of the turbulence equilibrium arguments of \cite{Zhang2012} for the logarithmic region. The transformation is an algebraic function that relates the local mean strain rate of the compressible flow, $d\widetilde{U}/dy$, to the non-dimensional incompressible mean strain-rate, $S_t^+$, at the same semi-local friction Reynolds number, $Re_\tau^*$, according to the relation 
\begin{equation} \label{eq:forward}
S_t^+ = \frac{S_{eq}^+}{1+S_{eq}^+-S_{TL}^+},
\end{equation}
where $S_{eq}^+=1/\overline{\mu}^+ d\widetilde{U}^+/dy^*$ and $S_{TL}^+=\overline{\mu}^+ d\widetilde{U}^+/dy^+$. The superscript $(\cdot)^*$ denotes non-dimensionalization by the local density $\rho(y)$, local molecular dynamic viscosity $\mu(y)$, and the semi-local friction velocity $u_{sl}=\sqrt{\tau_w/\overline{\rho}(y)}$ \citep{Huang1995,Coleman1995}. The semi-local friction Reynolds number is thus defined as $Re_\tau^* = \overline{\rho}_e u_{sl} \delta / \overline{\mu}_e$, where the subscript $(\cdot)_e$ denotes quantities evaluated at the boundary layer edge (throughout this work, $\delta$ denotes the channel half height or the boundary-layer thickness). Note that all variables of the form $S_{(\cdot)}^+$ represent different local non-dimensionalizations of the compressible strain rate, which were designed in prior works with the target of equaling the strain rate implied by the incompressible law of the wall. For example, although $S_{TL}^+$ is equivalent to the viscous stress, it is also a non-dimensionalization of the mean strain rate in a compressible flow. $S_{TL}^+$ will exactly recover the incompressible strain rate of a flow with the equivalent viscous stress as long as the compressible flow also obeys $\mu^+=1$. Additionally, note that the transformation in Eq.~(\ref{eq:forward}) assumes a constant stress layer in the buffer region of the boundary layer, where there is a transition between the underlying viscous and equilibrium transformations. \cite{Griffin2021a} verifies that the deployment of this assumption does not significantly affect the accuracy of the transformation in equilibrium flows, and \cite{Bai2022} verifies the same for boundary layers with moderate pressure gradients.

The inverse velocity transformation is readily obtained by algebraically rearranging the transformation to find
\begin{equation} \label{eq:inverse}
	\tder{\widetilde{U}^+}{y^*} = \left(\frac{1}{\overline{\mu}^+ S^+_{t}} - \frac{1}{\overline{\mu}^+}  +\sqrt{\overline{\rho}^+} \left(1 + \frac{1}{2\overline{\rho}^+}\tder{\overline{\rho}^+}{y^+} y^+ - \frac{1}{\overline{\mu}^+}\tder{\overline{\mu}^+}{y^+}y^+  \right) \right)^{-1}.
\end{equation}
The incompressible mean strain rate $S_t^+$ is available algebraically from the constant-property version of Eq.~(\ref{eq:u_ode}), i.e., $\overline{\rho}=\overline{\rho}_w$ and $\overline{\mu}=\overline{\mu}_w$. The incompressible model constants $\kappa$ and $B$ are determined using the aforementioned calibration but $Re_\tau^*$ is used in place of $Re_\tau$ since the former is invariant under the velocity transformation. Integrating Eq.~(\ref{eq:inverse}) with variable properties yields the targeted compressible velocity profile; the properties are functions of temperature, which will be discussed next.

\subsection{Algebraic temperature-velocity relation}
In order to close the velocity equation (Eq.~(\ref{eq:inverse})), the temperature profile must be determined. The classical model uses the constant turbulent Prandtl number assumption to develop a coupled ODE for temperature (Eq.~(\ref{eq:T_ode})). However, the constant Prandtl number assumption has been shown to be less accurate than invoking the Generalized Reynolds Analogy (GRA) \cite{Zhang2014}. Thus, the presently proposed wall model leverages the GRA instead.

The analogy between the conservation equations for momentum and energy has led to the derivation of several algebraic relations between temperature and velocity.
Walz's equation \citep{Walz1969} (also known as the modified Crocco-Busemann relation \citep{Crocco1932,Busemann1931})  leverages the analogy between the conservation equations for momentum and energy to arrive at an algebraic relation between mean temperature and velocity. This relation accounts for non-unity $Pr$ effects via a recovery factor, which is taken as $r=(\Pr)^{1/3}$. While this relation is accurate in high-speed adiabatic boundary layers, \cite{Duan2011_part4} observed that the accuracy degrades significantly in boundary layers with wall heat transfer and proposed a semi-empirical correction to the relation. 
This was subsequently recast in terms of a generalized Reynolds analogy \citep{Zhang2014}, thereby introducing the Reynolds analogy factor, $s$, which they choose as $s = 1.14$ following convention. The resulting temperature-velocity relation is given as,
\begin{equation} \label{eq:DM4}
\widetilde{T} = \widetilde{T}_w + s \Pr (\widetilde{T}_r-\widetilde{T}_w) \frac{\widetilde{U}}{\widetilde{U}_e} \left(1 - \frac{\widetilde{U}}{\widetilde{U}_e} \right) + \left( \frac{\widetilde{U}}{\widetilde{U}_e} \right)^2 \left( \widetilde{T}_e-\widetilde{T}_w \right),
\end{equation}
where the subscript $(\cdot)_e$ denotes quantities at the boundary-layer edge, the recovery temperature $\widetilde{T}_r = \widetilde{T}_e + r \widetilde{U}_e^2/(2 C_p)$. 
This relation has been validated across a wide range of channel flows, pipe flows, and boundary layers with and without heat transfer \citep{Zhang2014, Zhang2018, Volpiani2020a, Modesti2019, Fu2021}. Specifically, this relation is derived by \cite{Zhang2014} through defining the generalized recovery temperature $\widetilde{T}_{r_g} = \widetilde{T} + r_g \widetilde{U}^2/(2 C_p)$. Then, it is assumed that $\widetilde{T}_{r_g} = \widetilde{T}_w + U_s \widetilde{U}/C_p$,
where $U_s$ is a constant velocity scale. Equivalently, the assumption can be reinterpreted that $\widetilde{T}$ can be approximately represented as a second order Taylor expansion in terms of powers of $\widetilde{U}$, i.e.,
\begin{equation}
\widetilde{T} = b_0 + b_1 \widetilde{U} + b_2 \widetilde{U}^2/2,
\end{equation}
where the no-slip condition implies $b_0 = \widetilde{T}_w$, $b_1 = (\tderi{\widetilde{T}}{\widetilde{U}})|_w$. 
The algebraic relation of \cite{Zhang2014} can be recovered if $b_2$ is specified by evaluating the expression at the boundary-layer edge $\widetilde{T}_e=\widetilde{T}|_{\widetilde{U}_e}$ and $b_1$ is determined using the Reynolds analogy. However, in this work, we use the matching data (denoted with subscript $(\cdot)_m$) $\widetilde{T}_m=\widetilde{T}|_{\widetilde{U}_m}$ to set $b_2$, such that the exact value at the matching location can be enforced.
The final temperature-velocity relation is
\begin{equation} \label{eq:DM4_1}
\widetilde{T} = \widetilde{T}_w + s \Pr (\widetilde{T}_r-\widetilde{T}_w) \frac{\widetilde{U}}{\widetilde{U}_e} \left(1 - \frac{\widetilde{U}}{\widetilde{U}_m} \right) + \left( \frac{\widetilde{U}}{\widetilde{U}_m} \right)^2 \left( \widetilde{T}_m-\widetilde{T}_w \right).
\end{equation}
Note that one consequence of this relation is that the wall heat flux and wall shear stress are algebraically linked by the Reynolds analogy factor, where the heat flux is defined as $q_w = s \tau_w C_p (\widetilde{T}_w-\widetilde{T}_r)/\widetilde{U}_e $. 

\subsection{Implementation details} 
Like the classical model (Eqs.~(\ref{eq:u_ode}--\ref{eq:T_ode})), the present model requires a matching temperature, velocity, and density, an equation of state (the ideal gas law is used in this work and the thin-boundary-layer assumption implies the pressure is constant), and a viscosity law (either a power law or Sutherland's law depending on the relevant reference data). In addition, the present model requires as input the velocity and temperature at the boundary-layer edge (computed using the method of \cite{Griffin2021}) for deploying the algebraic temperature-velocity relation (Eq.~(\ref{eq:DM4_1})) due to its dependence on the recovery temperature and edge velocity. To solve the nonlinear system, the following approach is used. The incompressible ODE (Eq.~(\ref{eq:u_ode})) with constant properties is integrated once analytically, rearranged for $d\widetilde{U}/dy$ and substituted into the inverse velocity transformation (Eq.~(\ref{eq:inverse})) as $S$. This equation (initial value problem with an initial guess for the wall shear stress) is solved via the shooting method, where, at each integration step, a sub-iteration determines the velocity increment that is consistent with the temperature-velocity relation (Eq.~(\ref{eq:DM4_1})) and the resulting density and viscosity at that location. 

The implementation of the present model is available at the link provided in the data availability section at the end of this manuscript. This implementation was first developed by \cite{Griffin2021c} to compute temperature and velocity profiles for estimating grid-point requirements in compressible flows, and this manuscript serves as the comprehensive documentation and the further development of the underlying inverse method for WMLES approach for the first time. Intermediate developments were presented in \cite{Griffin2022_APS}, and initial results were reported in \cite{Griffin2022_ARB,Griffin2022_thesis}. \cite{Kumar2022} used a similar procedure but with a data-driven velocity transformation \citep{Volpiani2020}. \cite{chen2023linear} and \cite{Song2023} approximate the mean  profiles of channel flows by considering two velocity transformations \citep{Trettel2016,Griffin2021a} and employing the Central Mean Temperature Scaling \citep{song2022}.

\section{{\it A priori} results} \label{sec:a_priori}

\begin{figure}
	\centering
	\includegraphics[width=0.4\linewidth]{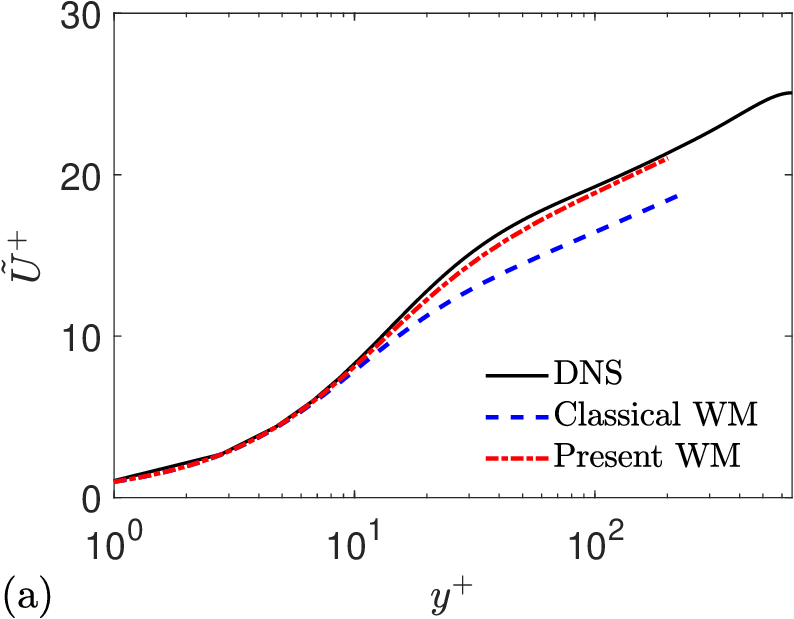}
	\includegraphics[width=0.415\linewidth]{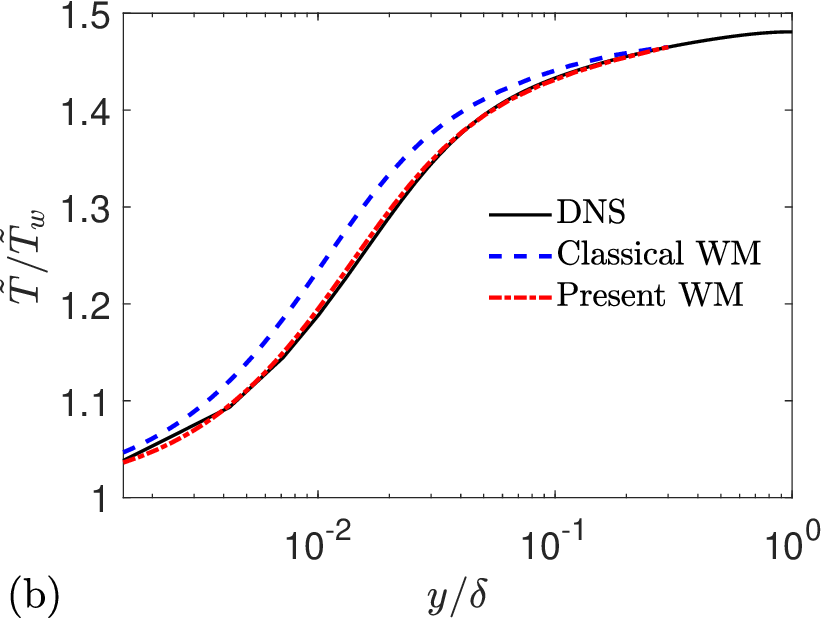}
	\includegraphics[width=0.4\linewidth]{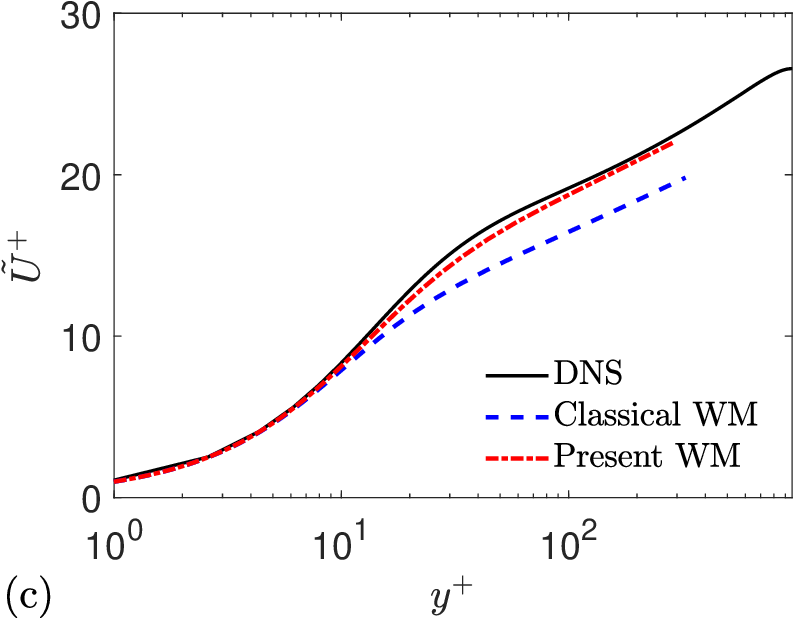}
	\includegraphics[width=0.4\linewidth]{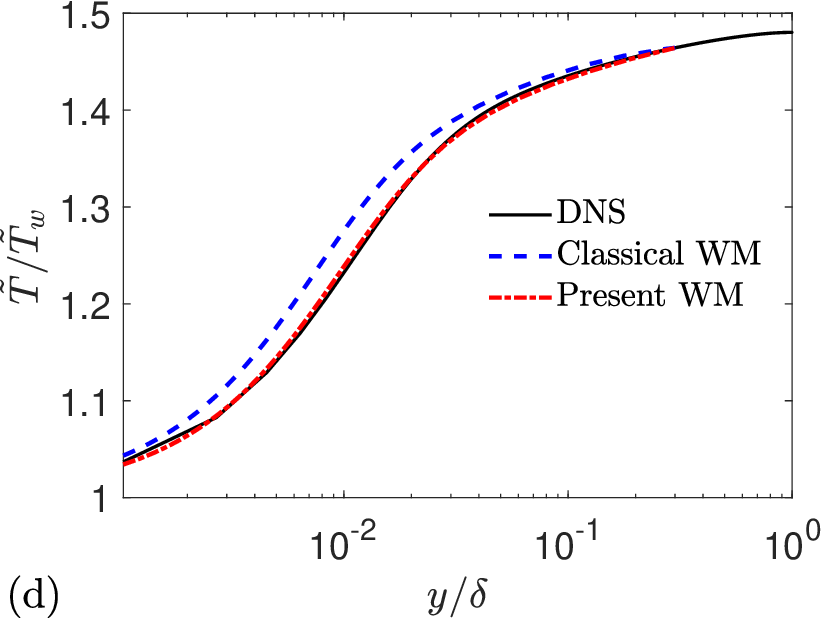}
	\caption{{\it A priori} wall-modeled profiles of velocity (a,c) and temperature (b,d) are plotted versus the wall-normal coordinate. Results are for supersonic channel flows with panels (a) and (b) characterized by $Re_\tau^* = 410$, $M_b = 1.7$, and $-B_q=0.053$ and panels (c) and (d) characterized by $Re_\tau^* = 590$, $M_b = 1.7$, and $-B_q=0.049$.}
	\label{fig:ap_profs_c1}
\end{figure}

\begin{figure}
	\centering
	\includegraphics[width=0.4\linewidth]{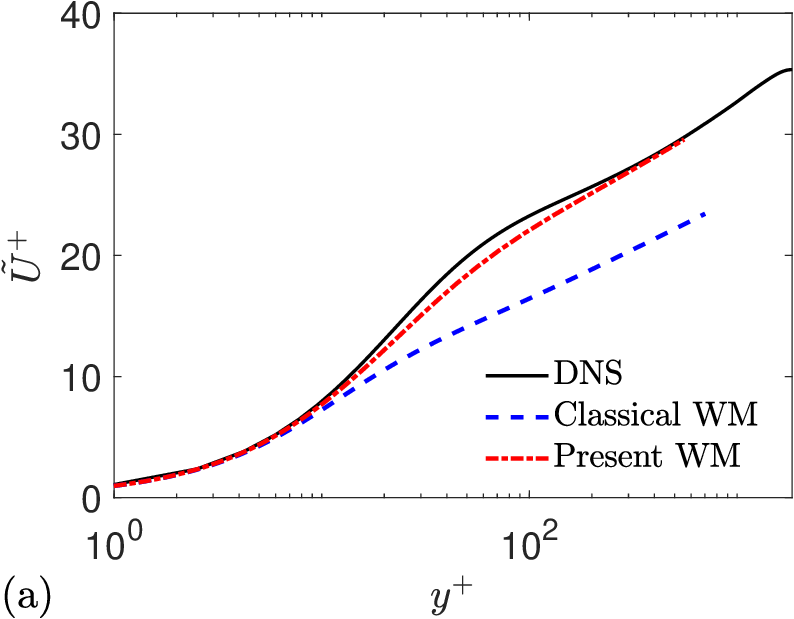}
	\includegraphics[width=0.415\linewidth]{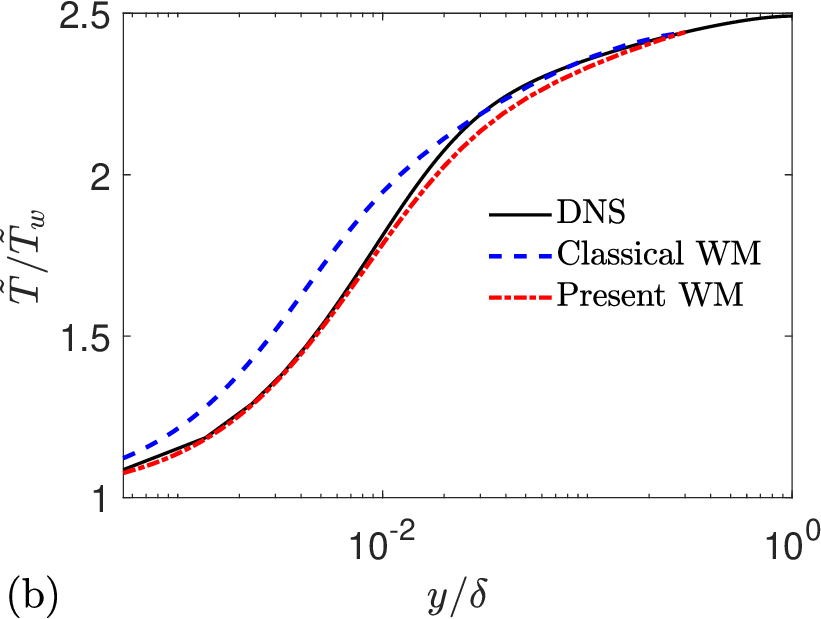}
	\includegraphics[width=0.4\linewidth]{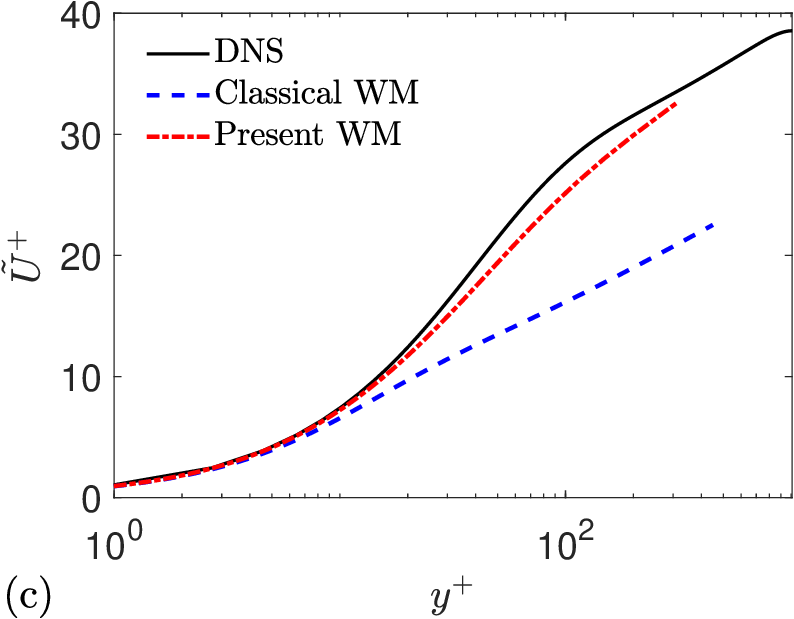}
	\includegraphics[width=0.4\linewidth]{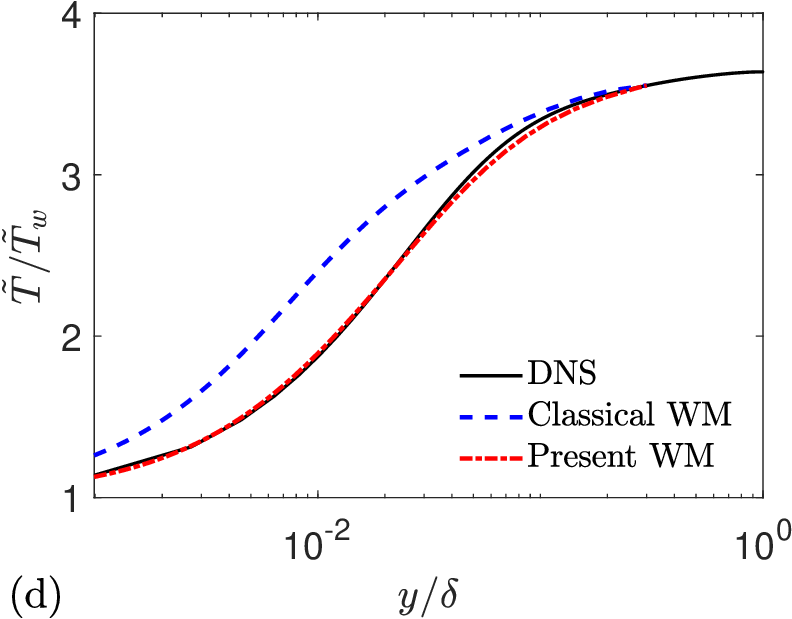}
	\caption{{\it A priori} wall-modeled profiles of velocity (a,c) and temperature (b,d) are plotted versus the wall-normal coordinate. Results are for supersonic channel flows with panels (a) and (b) characterized by $Re_\tau^* = 590$, $M_b = 3.0$, and $-B_q=0.12$ and panels (c) and (d) characterized by $Re_\tau^* = 200$, $M_b = 4.0$, and $-B_q=0.19$.}
	\label{fig:ap_profs_c2}
\end{figure}

\begin{figure}
	\centering
	\includegraphics[width=0.4\linewidth]{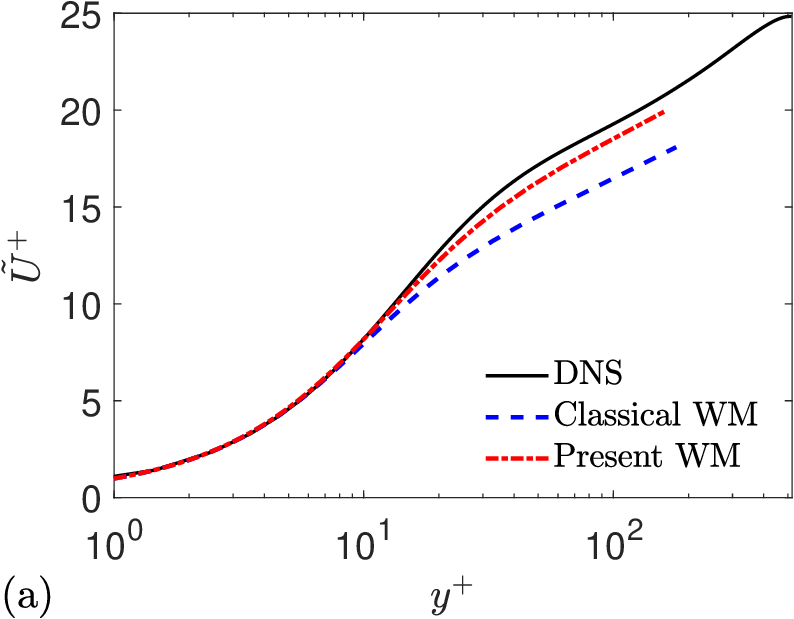}
	\includegraphics[width=0.415\linewidth]{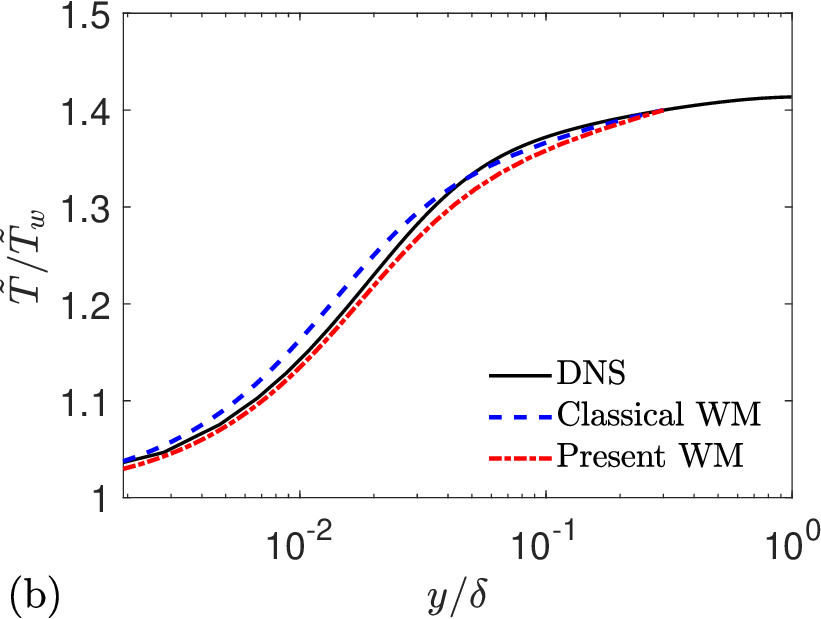}
	\includegraphics[width=0.4\linewidth]{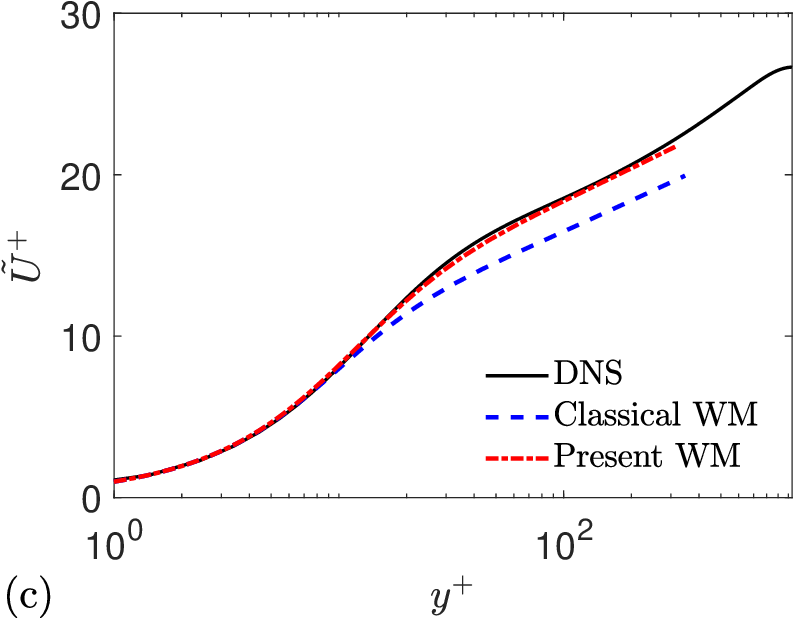}
	\includegraphics[width=0.4\linewidth]{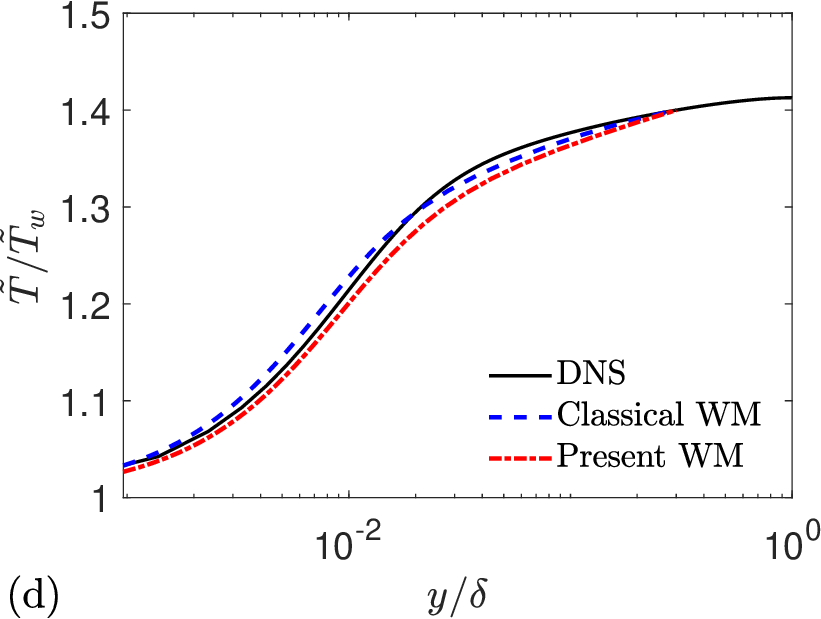}
	\caption{{\it A priori} wall-modeled profiles of velocity (a,c) and temperature (b,d) are plotted versus the wall-normal coordinate. Results are for supersonic pipe flows with panels (a) and (b) characterized by $Re_\tau^* = 333.5$, $M_b = 1.500$, and $-B_q=0.047$ and panels (c) and (d) characterized by $Re_\tau^* = 668.8$, $M_b = 1.500$, and $-B_q=0.044$.}
	\label{fig:ap_profs_p}
\end{figure}

\begin{figure}
	\centering
	\includegraphics[width=0.4\linewidth]{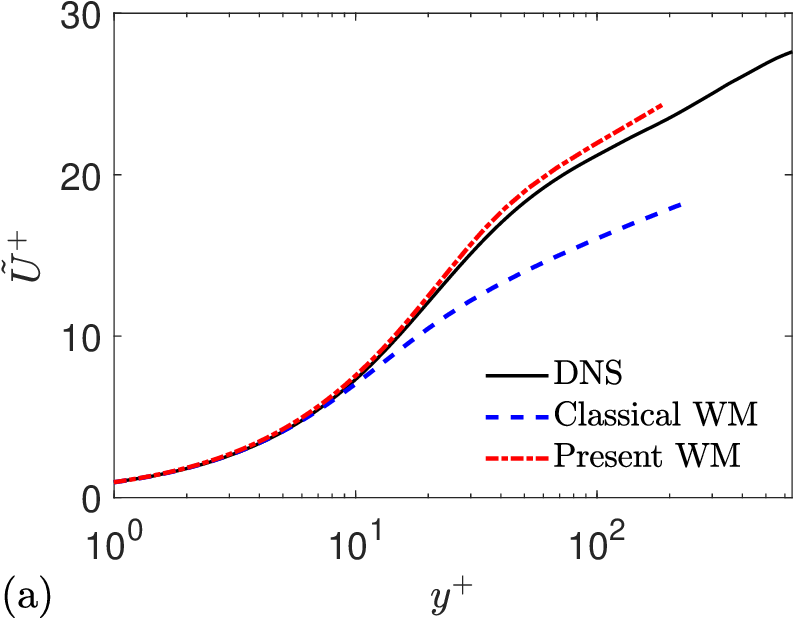}
	\includegraphics[width=0.415\linewidth]{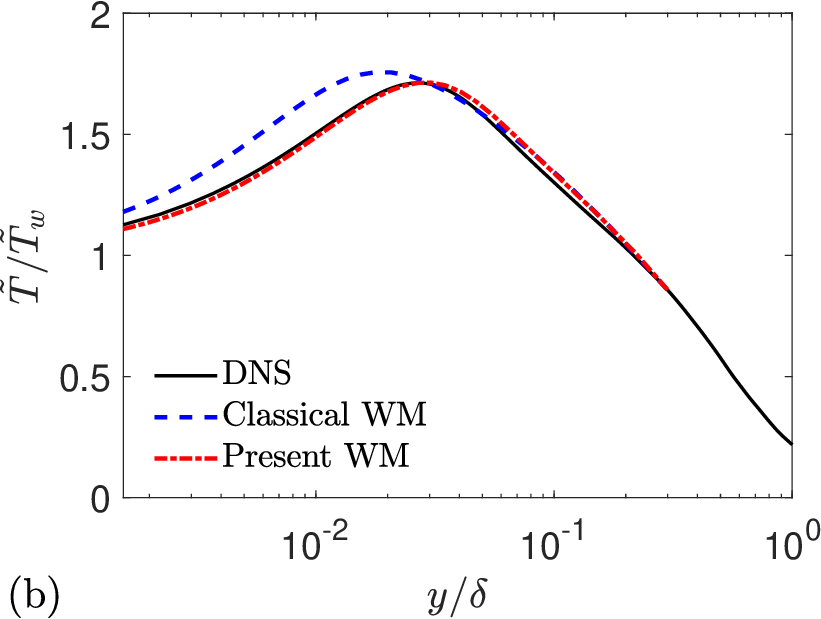}
	\includegraphics[width=0.4\linewidth]{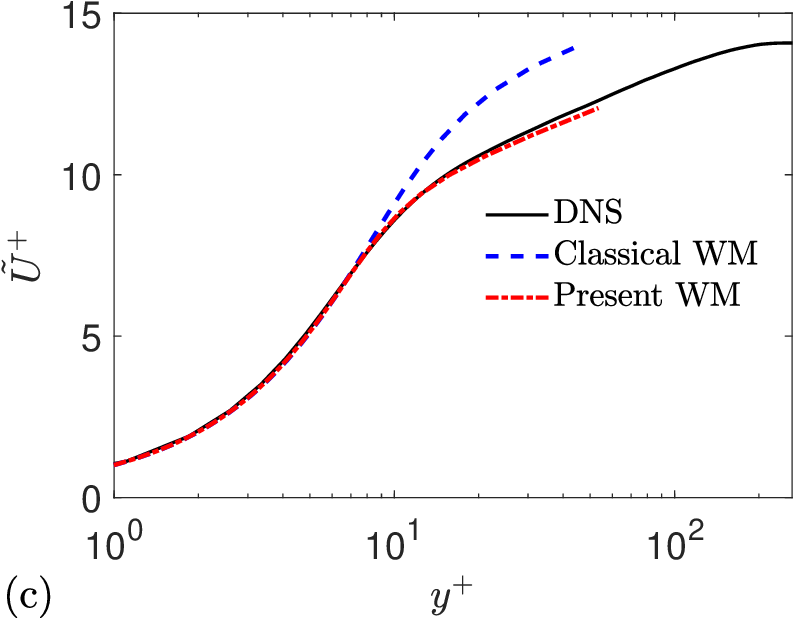}
	\includegraphics[width=0.4\linewidth]{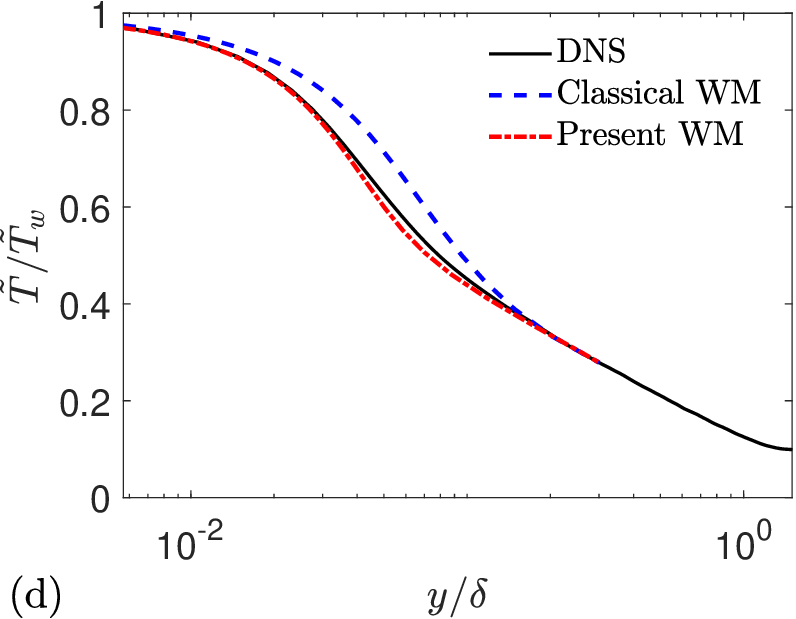}
	\caption{{\it A priori} wall-modeled profiles of velocity (a,c) and temperature (b,d) are plotted versus the wall-normal coordinate. Results are for hypersonic diabatic (isothermal) boundary layers with panels (a) and (b) characterized by $Re_\tau^* = 5677$, $M_e = 11.46$, and $-B_q=0.19$ (cooled wall) and panels (c) and (d) characterized by $Re_\tau^* = 2328$, $M_e = 4.327$, and $-B_q=-0.039$ (heated wall).}
	\label{fig:ap_profs_bl}
\end{figure}

The present and classical wall models are first evaluated via {\it a priori} analysis. That is, the matching data are taken from DNS at a wall-normal distance of $y_m=0.3\delta$. The wall model estimates the velocity and temperature profiles, as well as the wall shear stress and wall heat flux. The predicted velocity and temperature profiles are shown in Figure \ref{fig:ap_profs_c1} and \ref{fig:ap_profs_c2} for four channel flows with various Mach and Reynolds number conditions, Figure \ref{fig:ap_profs_p} for two pipe flows at different Reynolds numbers, and Figure \ref{fig:ap_profs_bl} for two boundary layers, one with a heated and one with a cooled wall boundary condition. The bulk Mach number is defined as $M_b = U_b/\sqrt (\gamma R \widetilde{T}_w)$, where $\gamma$ is the ratio of specific heats and $R$ is the gas constant. The bulk Reynolds number is defined as $Re_b = \rho_b U_b \delta / \overline{\mu}_w$, where the bulk density is defined as $\rho_b = \iint_A \overline{\rho} dA/A$ and the bulk velocity is defined as $U_b = \iint_A \widetilde{U} dA/A$, where $A$ is the cross-sectional area of the domain. Reference DNS data are provided by \cite{Modesti2019,Trettel2016,Zhang2018,Volpiani2020a}.
For all cases, the profiles predicted by the present model agree with the DNS profiles significantly better than the classical model. Note that the velocities are non-dimensionalized by the predicted friction velocity, so the obtained profiles do not necessarily pass through the matching data if the predicted wall stress is inaccurate.

\begin{figure}
	\centerline{
	\includegraphics[width=0.4\linewidth]{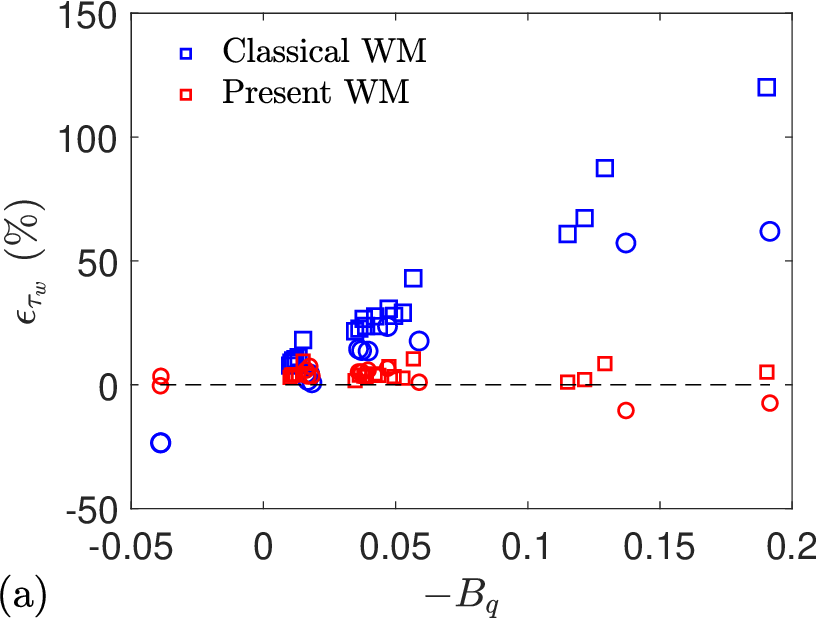}
	\includegraphics[width=0.4\linewidth]{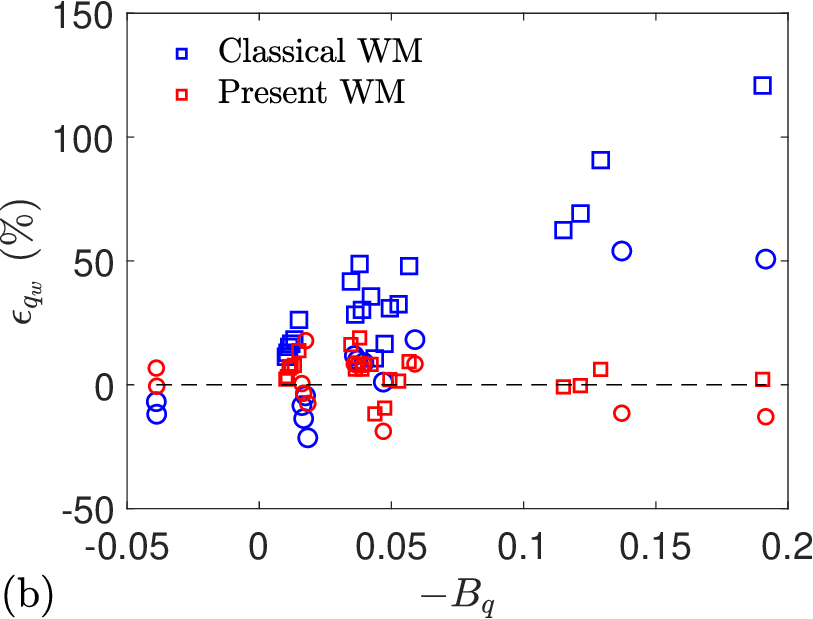}}
	\caption{{\it A priori} modeling errors of the wall shear stress $\tau_w$ (a) and the wall heat flux $q_w$ (b) versus the heat transfer coefficient $B_q$. The model matching data are taken from DNSs of various channel and pipe flows (squares), nearly adiabatic boundary layers (triangles), and diabatic boundary layers (circles). 
	}
	\label{fig:ap_sweep}
\end{figure}

Next, the model performance is evaluated with a wide range of DNS data from 48 different simulations. 
The errors in the modeled wall stress and heat flux predictions are reported for each case with $y_m=0.3\delta$. The relative error in the wall stress prediction $\epsilon_{\tau_w}$ is defined as
\begin{equation} \label{eq:tauw}
	\epsilon_{\tau_w} =  \frac{\tau_{w,\mathrm{model}} - \tau_{w,\mathrm{DNS}}}{\tau_{w,\mathrm{DNS}}} \times 100\%.
\end{equation}
The non-dimensional wall heat flux is defined as $B_q = q_{w}/(C_p \widetilde{T}_w \overline{\rho}_w u_{\tau})$, and the relative error in the wall heat flux is defined as
\begin{equation} \label{eq:qw}
	\epsilon_{q_w} = \frac{ q_{w,\mathrm{model}} - q_{w,\mathrm{DNS}} }{q_{w,\mathrm{DNS}}}  \times 100\%.
\end{equation}
$\epsilon_{q_w}$ is not reported for adiabatic boundary layer data because it is undefined, and both models predict negligible heat transfer for these data.
The data considered include the compressible channel flow simulations of \cite{Modesti2016,Trettel2016,Yao2020}, the pipe flow simulations of \cite{Modesti2019}, the adiabatic supersonic and hypersonic boundary layers of \cite{Pirozzoli2011,Zhang2018,Volpiani2018,Volpiani2020a}, and the diabatic supersonic and hypersonic boundary layers of \cite{Zhang2018,Volpiani2018,Volpiani2020a,Fu2019}. 
The cases have edge Mach numbers in the range of 0.77--11 and semi-local friction Reynolds numbers in the range of 170--5700.
Only the cases with $Re_\tau^* > 150$ are analyzed because lower Reynolds numbers can exhibit strong Reynolds number effects \citep{Modesti2016} and are not the target of this study. The error measures are shown in Figure \ref{fig:ap_sweep}. The present model generates significantly less modeling error than the classical model, with the greatest error reduction when the non-dimensional heat transfer is the highest.
%
\begin{figure}
	\centering
	\includegraphics[width=0.4\linewidth]{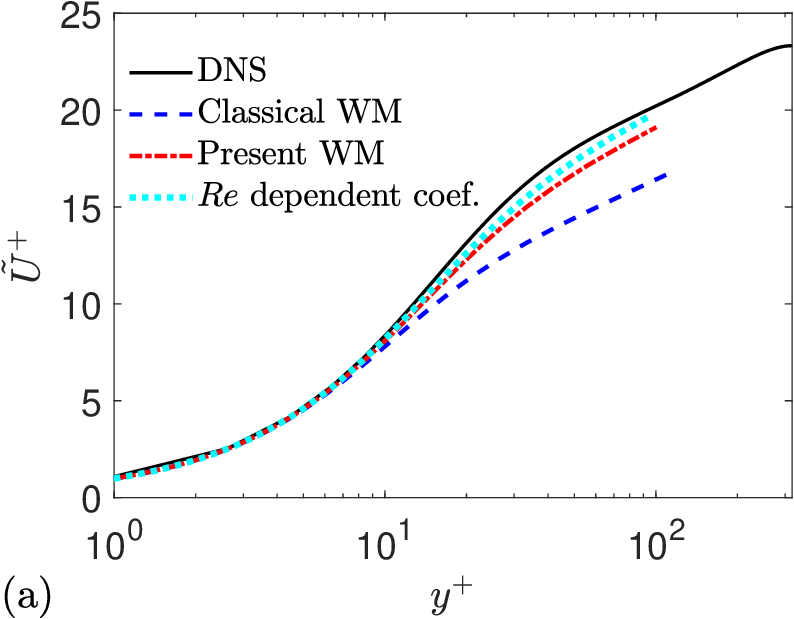}
	\includegraphics[width=0.41\linewidth]{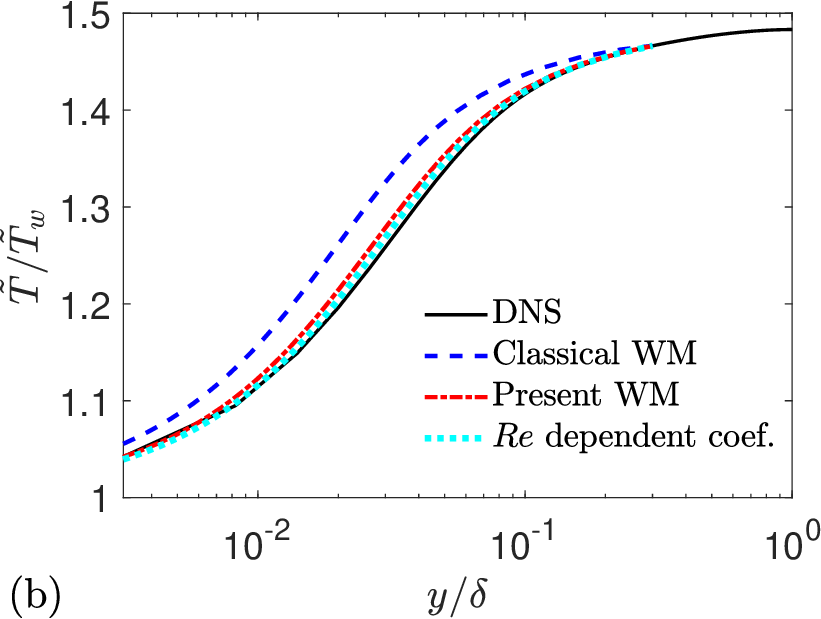}
    \includegraphics[width=0.4\linewidth]{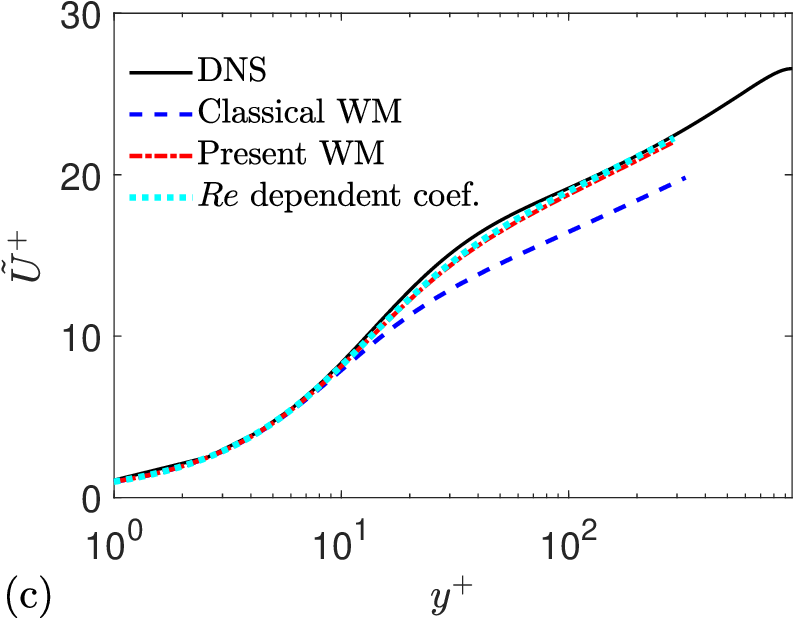}
	\includegraphics[width=0.41\linewidth]{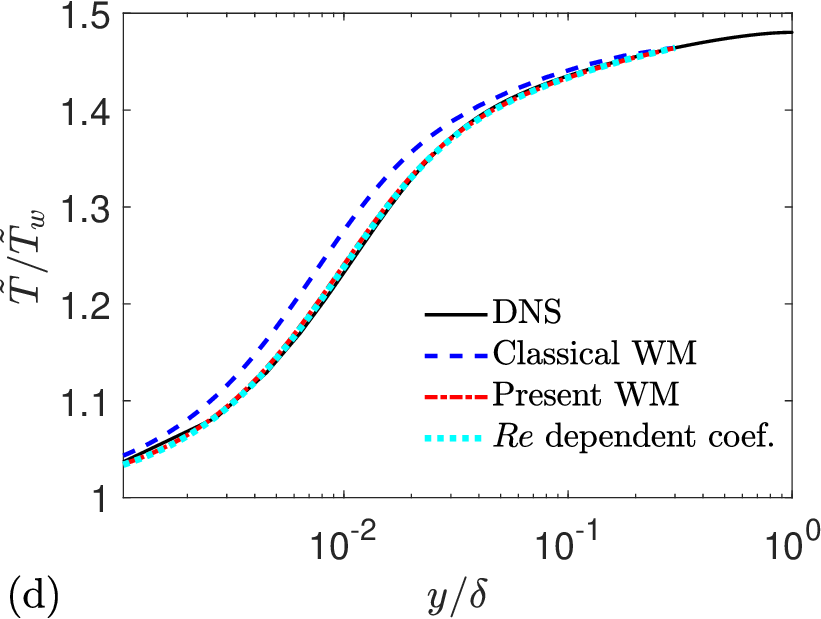}
	\caption{{\it A priori} wall-modeled profiles of velocity (a,c) and temperature (b,d) are plotted versus the wall-normal coordinate. Results are for supersonic channel flows with panels (a) and (b) characterized by $Re_\tau^* = 190$, $M_b = 1.7$, and $-B_q=0.057$ and panels (c) and (d) characterized by $Re_\tau^* = 590$, $M_b = 1.7$, and $-B_q=0.049$.}
	\label{fig:ap_profs_ode_coef}
\end{figure}

To distinguish the effects of Reynolds number and compressibility, we explore the effect of using Reynolds-number-dependent coefficients for the underlying incompressible Law of the Wall. Specifically, rather than letting the von K{\'a}rm{\'a}n constant $\kappa$ and the damping coefficient $A^+$ be fixed values of 4.1 and 17, respectively, we recalibrate these values using incompressible reference data at various Reynolds numbers. We employ the DNS data from five incompressible turbulent channel flows \citep{Lee2015} with friction Reynolds numbers $Re_\tau = u_\tau \delta / \nu_w = \{182, 543, 1000, 1990, 5190\}$, and fit the least-squares optimal values of $\kappa = \{0.400, 0.408, 0.400, 0.391, 0.391\}$ and $A^+ = \{18.2, 17.4, 17.0, 16.5, 16.5\}$. Linear interpolation and constant extrapolation of the optimal values are used to define $\kappa$ and $A^+$ for all Reynolds numbers. The inverse velocity transformation uses the semi-local wall-normal coordinate $y^*$, so the incompressible data should be interpreted as a function of $Re_\tau^*$ rather than $Re_\tau$. {\it A priori} analysis is performed as before using compressible DNS data, but with the optimal coefficients selected according to the $Re_\tau^*$ observed in the compressible DNS. In Figure \ref{fig:ap_profs_ode_coef}(a-b), for the case of a turbulent channel flow with $Re_\tau^* = 190$ and $M_b = 1.7$, there is a modest improvement from using the Reynolds-number-dependent coefficients for the incompressible model. This suggests that at low Reynolds numbers, the deviation of DNS data for the incompressible constant-property velocity profile from the nominal law of the wall is on the same order as the deviation of the constant coefficient model and compressible DNS velocity profile. However, there is not a complete collapse of the model with Reynolds-number-dependent coefficients with the compressible DNS. This is likely attributed to the documented error in the compressible velocity transformation at $Re_\tau^* <\sim 200$ \cite{Griffin2021a}. In Figure \ref{fig:ap_profs_ode_coef}(c-d), the case of a turbulent channel flow with $Re_\tau^* = 590$ and $M_b = 1.7$ is considered. The Reynolds number is high enough that the optimal and constant coefficients are similar; thus, the performance of the present model with either set of coefficients is similar. Overall, there is no significant sensitivity to tuning the coefficients, so, for simplicity, we use the constant coefficients of $\kappa=0.41$ and $A^+=17$ for the remainder of this manuscript.

\begin{figure}
	\centering
	\includegraphics[width=0.4\linewidth]{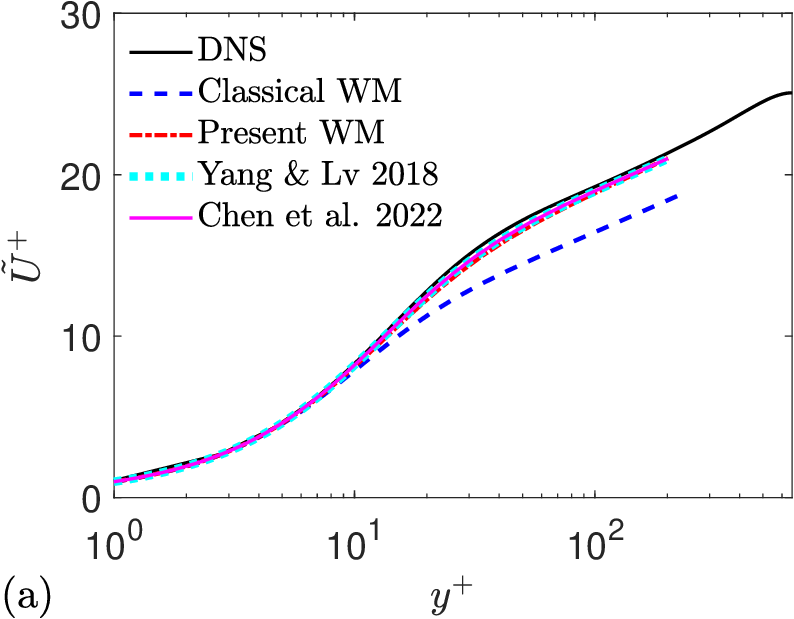}
	\includegraphics[width=0.41\linewidth]{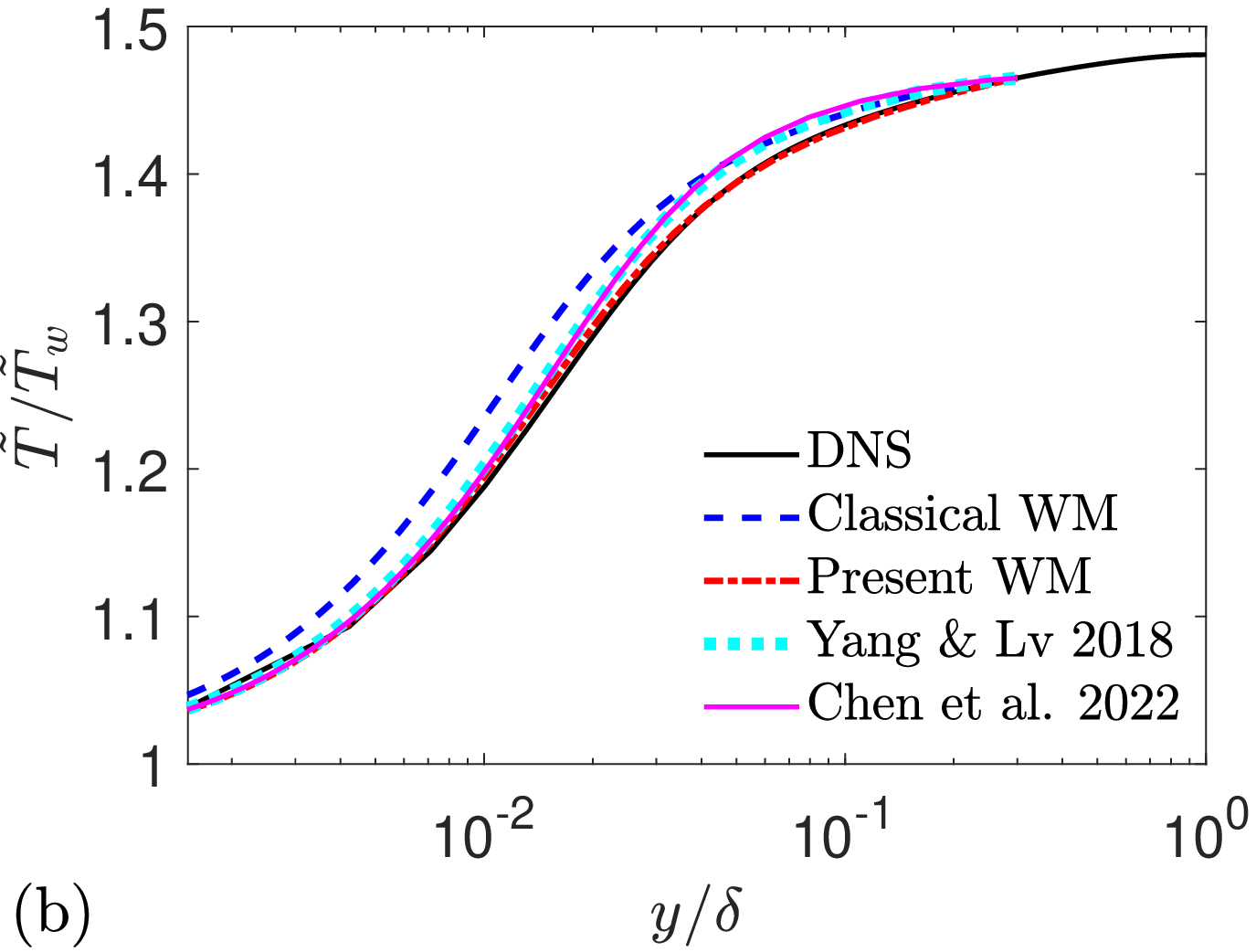}
    \includegraphics[width=0.4\linewidth]{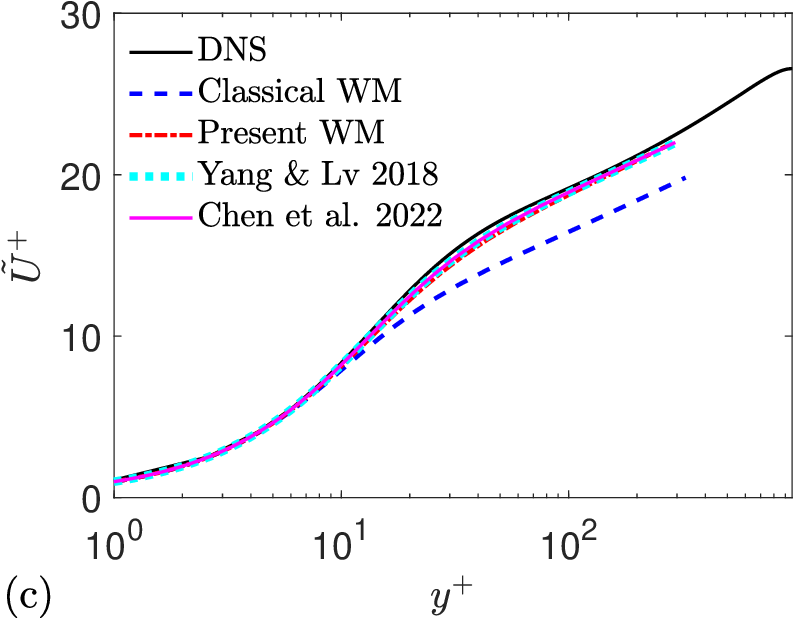}
	\includegraphics[width=0.41\linewidth]{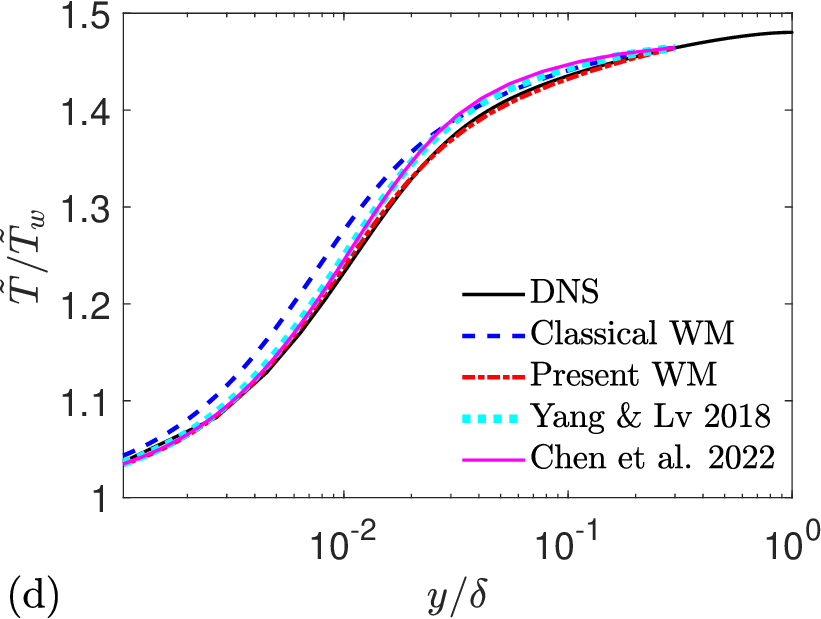}
    \includegraphics[width=0.4\linewidth]{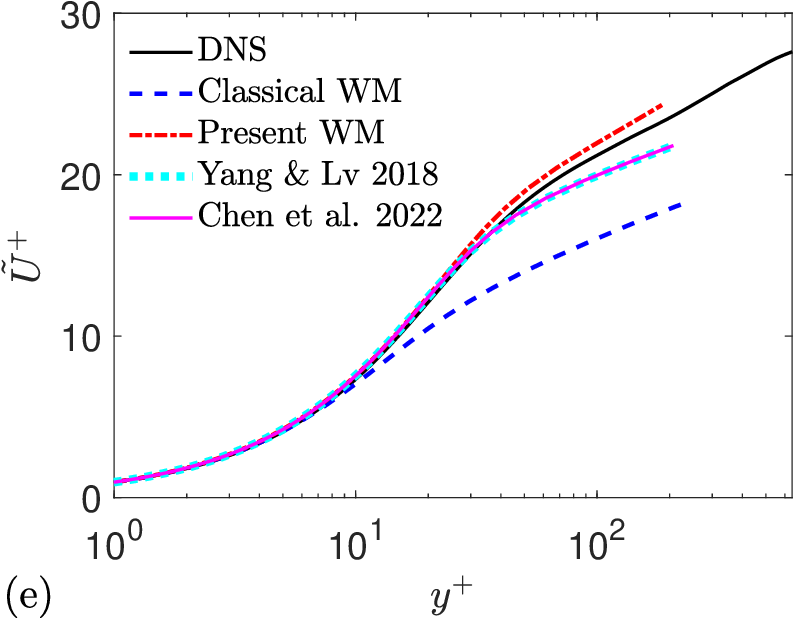}
	\includegraphics[width=0.41\linewidth]{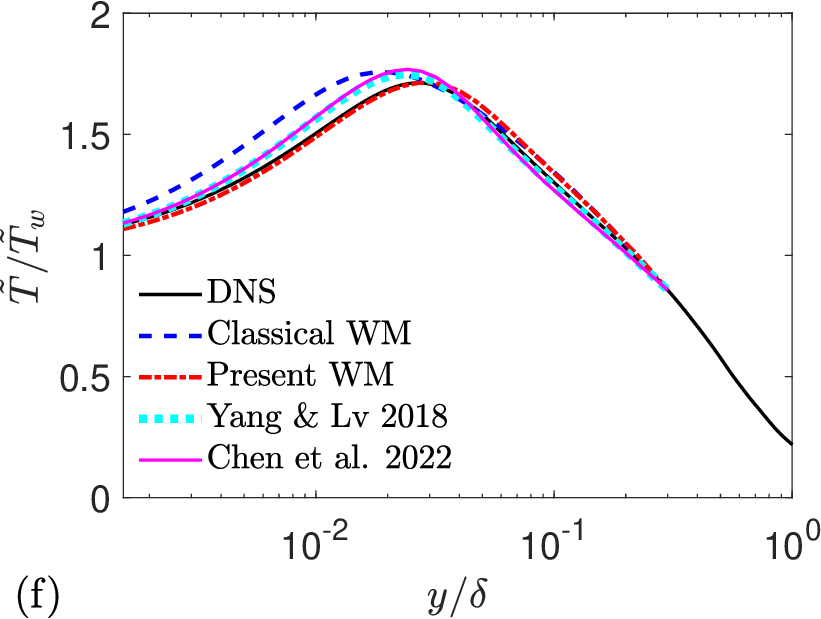}
    \includegraphics[width=0.4\linewidth]{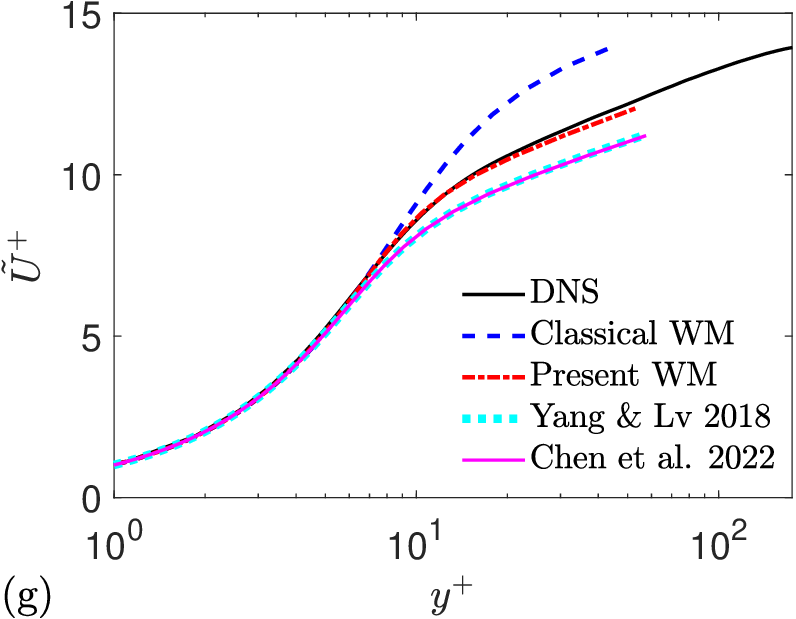}
	\includegraphics[width=0.41\linewidth]{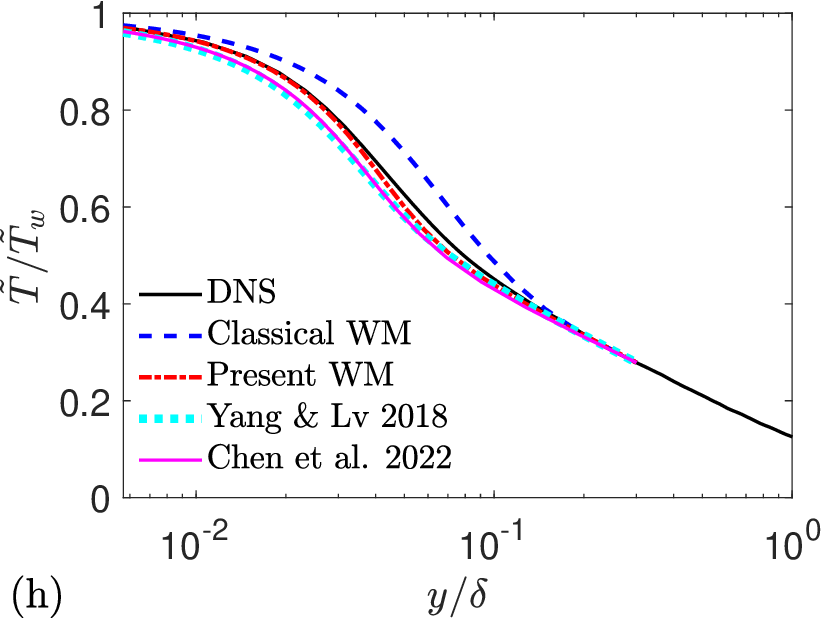}
	\caption{{\it A priori} wall-modeled profiles of velocity (a,c,e,g) and temperature (b,d,f,h) are plotted versus the wall-normal coordinate for four wall models. Panels (a-d) correspond to the supersonic channel flows presented in Figure \ref{fig:ap_profs_c1}; panels (e-h) correspond to the hypersonic diabatic boundary layers presented in Figure \ref{fig:ap_profs_bl}.}
	\label{fig:ap_profs_mods}
\end{figure}

Two more recently developed compressible wall models are considered. The first is developed by \cite{Yang2018a}; they show that the damping function in the classical model (Eq.~(\ref{eq:mu_t})) is consistent with the velocity transformation of \cite{VanDriest1951}, which has been shown to be less accurate in channel flows than the velocity transformation of \cite{Trettel2016}. Therefore, \cite{Yang2018a} rewrite the damping function in terms of $y^*$ and show that this makes the model consistent with the Trettel-Larsson transformation. The second additional model considered is proposed by \cite{Chen2022}, which also uses the semi-local damping function and further replaces the constant turbulent Prandtl number assumption of the classical model with an explicit function of $y^*$. In Figure \ref{fig:ap_profs_mods}, these two additional wall models are compared with the classical and present wall models. Figure \ref{fig:ap_profs_mods}(a-d) indicate that all models are performing well in the channel flows except for the classical model. This behavior is explained by the behavior of the underlying velocity transformations. The models of \cite{Yang2018a} and \cite{Chen2022} use the Trettel-Larsson transformation and the present model uses the total-stress-based transformation \citep{Griffin2021a}. Both of these transformations are well established to outperform the van Driest transformation (used by the classical model) in channel flows. In Figures \ref{fig:ap_profs_mods}(e-f) and \ref{fig:ap_profs_mods}(g-h), the models are applied to boundary layers with cooled and heated walls, respectively. For both cases the classical model is the least accurate likely due to the inaccuracy of the van Driest transformation for boundary layers with strong heat transfer \citep{Griffin2021a}, as the velocity transformation is the only difference between the classical model and that of \cite{Yang2018a}. Also for both cases, the models that use semi-local damping \citep{Yang2018a,Chen2022} perform almost identically, suggesting limited sensitivity in these flows to the change in turbulent Prandtl number model proposed by \cite{Chen2022}. For the heated boundary layer, the present model slightly improves the prediction of the temperature peak and the log slope of the velocity compared to the semi-local damping models. For the cooled boundary layer, there is a more substantial improvement from the present model for the log slope of the velocity but the temperature profiles are only slightly improved. These improvements of the present model over the semi-local damping models are consistent with the improvements of the total-stress-based transformation over the Trettel-Larsson transformation for boundary layers with strong heat transfer.
%

\section{{\it A posteriori} WMLES results} \label{sec:a_posteriori}

In this section, several WMLES simulations are conducted using charLES, a high-fidelity compressible finite-volume code \citep{Bres2018}. The numerical method consists of a low-dissipation, approximately entropy-preserving scheme, which utilizes artificial bulk viscosity to capture the solution discontinuities. Additional details about the solver and a summary of validation campaigns are available in \cite{Fu2021,Fu2022}.

The WMLESs conducted herein are compressible turbulent channel flows driven with uniform volumetric momentum and energy source terms to achieve the same bulk Mach number $M_b$ and bulk Reynolds number $Re_b$ conditions of the DNS simulations of \cite{Trettel2016} as summarized in table \ref{tab:cases}.

%
\begin{table}
	\begin{center}
		\def~{\hphantom{0}}
		\begin{tabular}{l||c|c|c|c|c|c|c|c|c}
			$M_b$       & 0.6998  & 0.6999  & 1.698   & 1.699   & 1.699   & 2.994  & 2.996   & 2.997  & 3.993 \\ 
$Re_b$      & 7498    & 11750   & 4495    & 9993    & 15490   & 7486   & 14980   & 23980  & 9979 \\
$Re_\tau$   & 436.6   & 650.9   & 318.6   & 661.6   & 963.6   & 636.4  & 1208    & 1842   & 1010.  \\
$Re_\tau^*$ & 395.7   & 590.0   & 194.8   & 405.4   & 590.8   & 204.0  & 387.7   & 589.7  & 201.3   \\
$-100 B_q$      & 1.061 & 1.009 & 5.668 & 5.273 & 4.942 & 12.92 & 12.15  & 11.50 & 19.04  \\
		\end{tabular}
		\caption{Non-dimensional flow parameters for the nine compressible turbulent channel flow cases considered for {\it a posteriori} testing within the WMLES framework.}
		\label{tab:cases}
	\end{center}
\end{table}

The cases are run on a domain of size $(\pi \times 2 \times \pi\sqrt{3}/4)\delta$ with periodic boundary conditions in the streamwise (first) and spanwise (third) dimensions. The mean profiles and fluxes were insensitive to doubling of the streamwise and spanwise domain sizes. Consistent with the DNS simulations, the viscosity is described by $\mu/\mu_{ref}=(T/T_{ref})^{0.75}$ and $Pr = 0.7$. All cases are initialized from a uniform solution with the target bulk Mach number and Reynolds number, and zero velocity in the wall-normal and spanwise directions. The simulations are allowed to transition from laminar to turbulent states naturally and are run for $\sim500$ eddy turnover times $\delta/u_\tau$. To challenge the wall model and isolate the effect of near-wall numerical errors \citep{Kawai2012}, the wall model matching location is placed at $y_m=0.3\delta$ and a coarse grid of 12 points per half channel height is used for all simulations unless otherwise indicated. The computational cost of the present model is similar to that of the classical model. The present model varies between being 7\% faster and 32\% slower depending on the Reynolds number, matching location, and Mach number. No effort was made to optimize the performance of the present model, so these numbers are just meant to indicate that the approximate cost of the model is similar in the cases tested. In general, modest differences in the cost of a wall model can be efficiently amortized over parallel processors via load balancing that assigns fewer control volumes to processors that contain more boundary faces, but this is not used in the present study.

\begin{figure}
	\centering
	\includegraphics[width=0.4\linewidth]{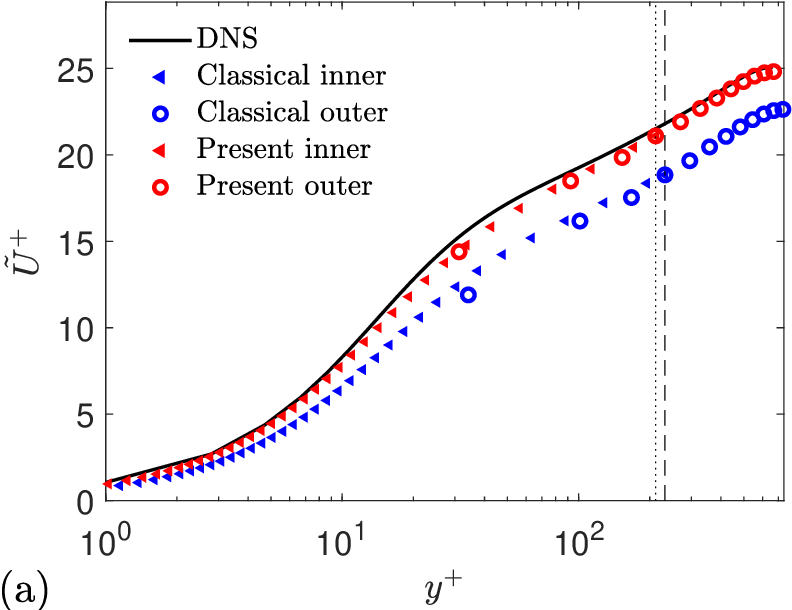}
	\includegraphics[width=0.405\linewidth]{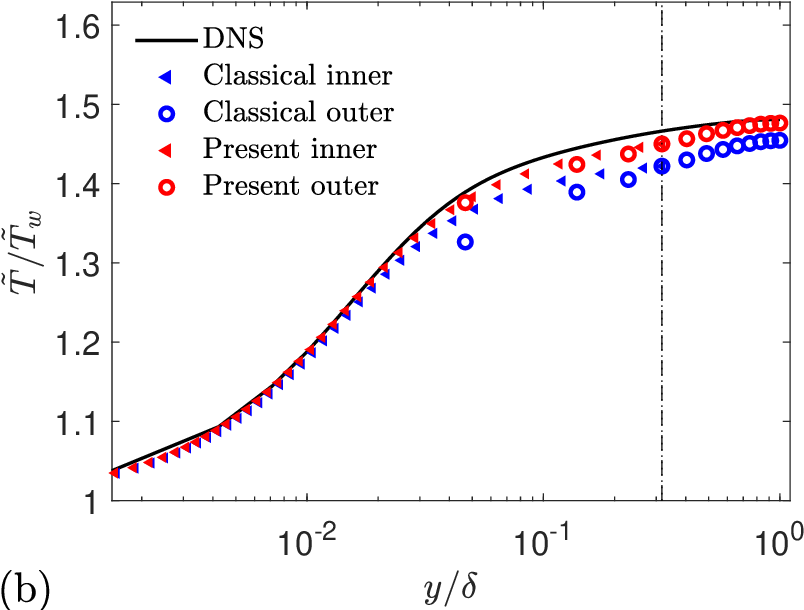}
	\\
	\includegraphics[width=0.4\linewidth]{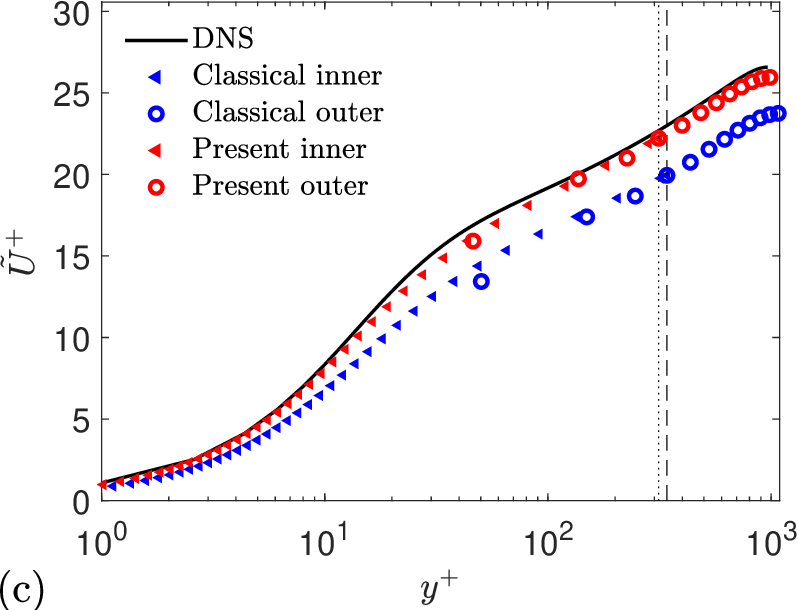}
	\includegraphics[width=0.405\linewidth]{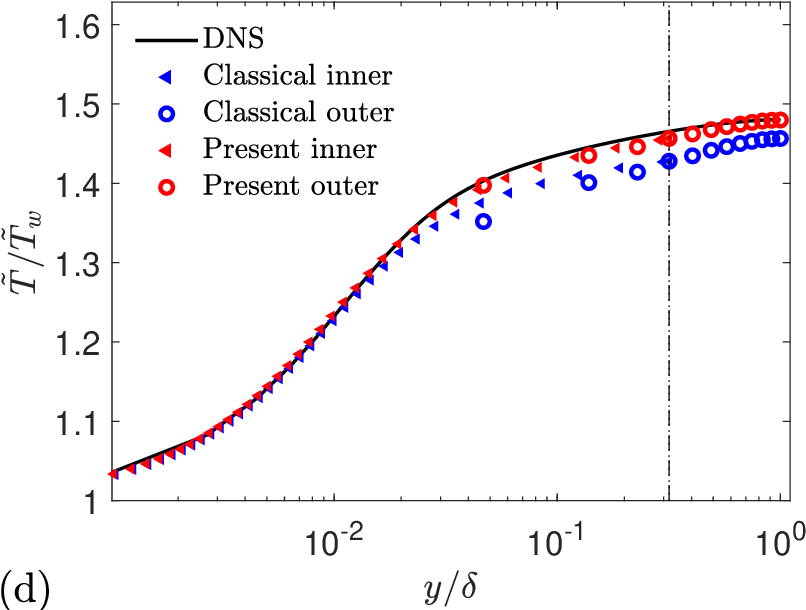}
	\caption{Velocity (a,c) and temperature (b,d) profiles from WMLES with the classical (blue) and present (red) wall models. A channel flow with $M_b=1.7$, $Re_\tau^*=410$, and $-B_q = 0.053$ is shown in panels (a) and (b), and one with $M_b=1.7$, $Re_\tau^*=590$, and $-B_q = 0.049$ is shown in panels (c) and (d). Within the WMLES framework, the outer solutions are computed by the LES PDE solver, while the inner solutions are computed by the two wall models. These solutions coincide at the LES matching point nearest to $y_m=0.3\delta$, which is indicated with the dashed and dotted lines for the classical and present models, respectively.}
	\label{fig:les_profs_M2}
\end{figure}

\begin{figure}
	\centering
	\includegraphics[width=0.4\linewidth]{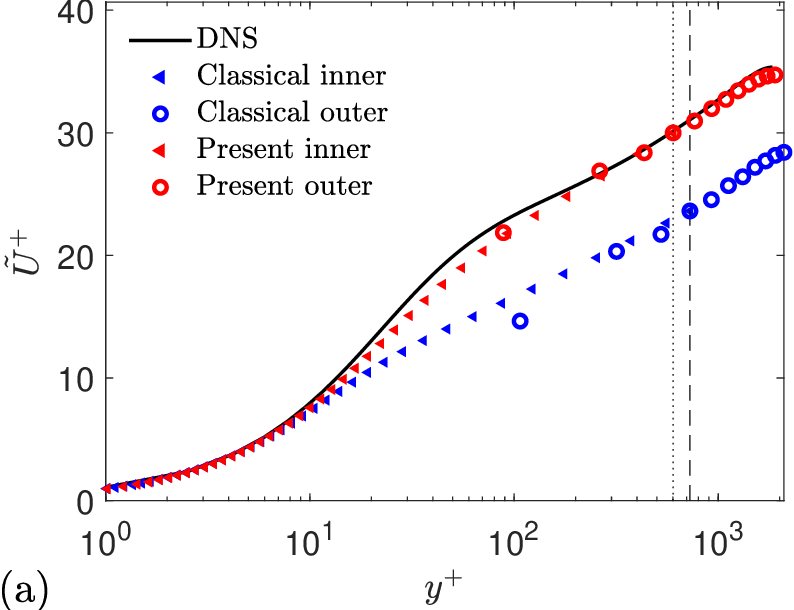}
	\includegraphics[width=0.405\linewidth]{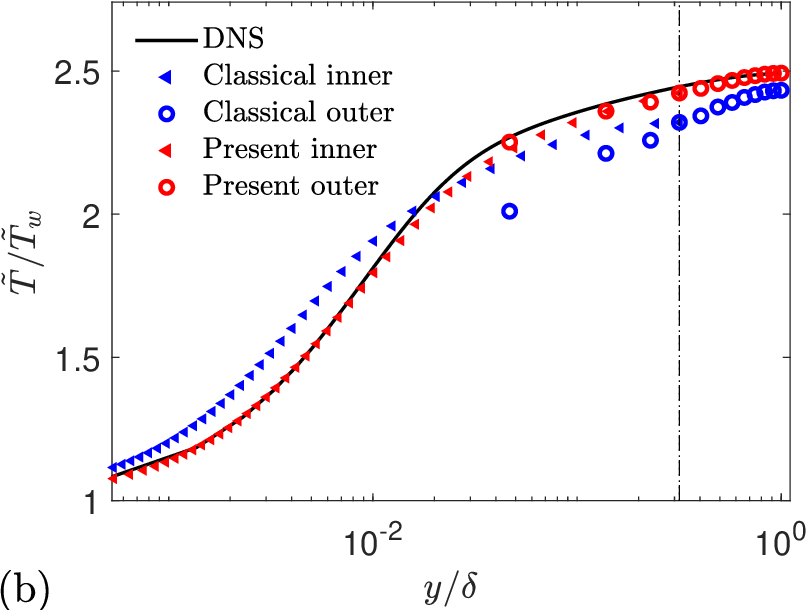}
	\\
	\includegraphics[width=0.4\linewidth]{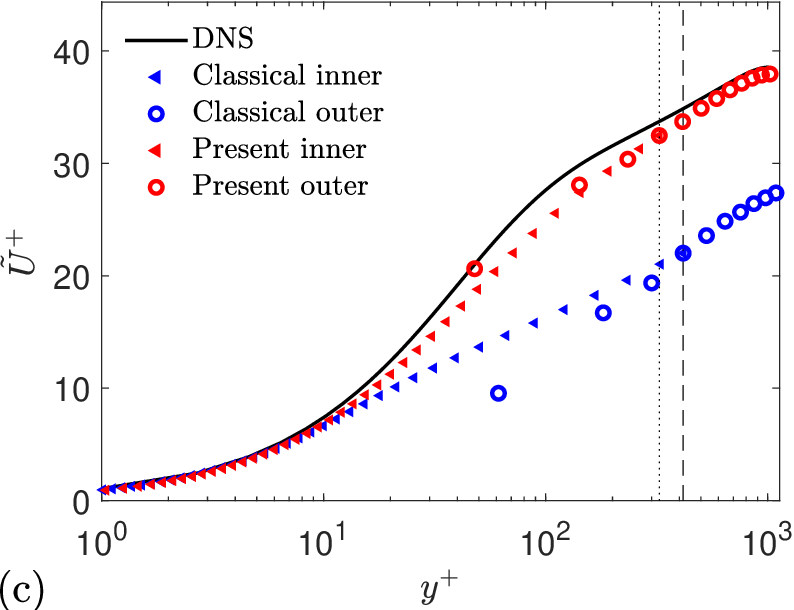}
	\includegraphics[width=0.405\linewidth]{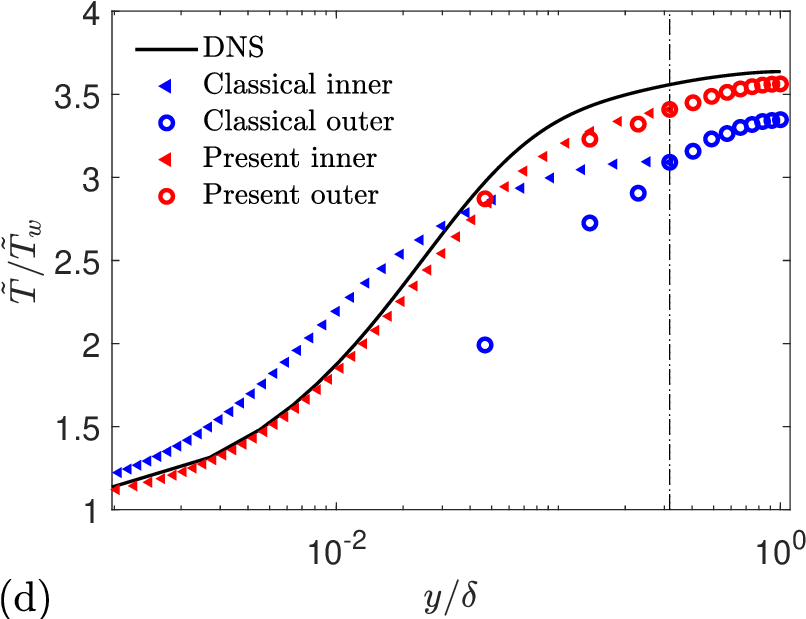}
	\caption{Velocity (a,c) and temperature (b,d) profiles from WMLES with the classical (blue) and present (red) wall models. A channel flow with $M_b=3.0$, $Re_\tau^*=590$, and $-B_q=0.12$ is shown in panels (a) and (b), and one with $M_b=4.0$, $Re_\tau^*=200$, and $-B_q=0.19$ is shown in panels (c) and (d). Within the WMLES framework, the outer solutions are computed by the LES PDE solver, while the inner solutions are computed by the two wall models. These solutions coincide at the LES matching point nearest to $y_m=0.3\delta$, which is indicated with the dashed and dotted lines for the classical and present models, respectively.}
	\label{fig:les_profs}
\end{figure}

\begin{figure}
	\centerline{
		\includegraphics[height=0.3\linewidth]{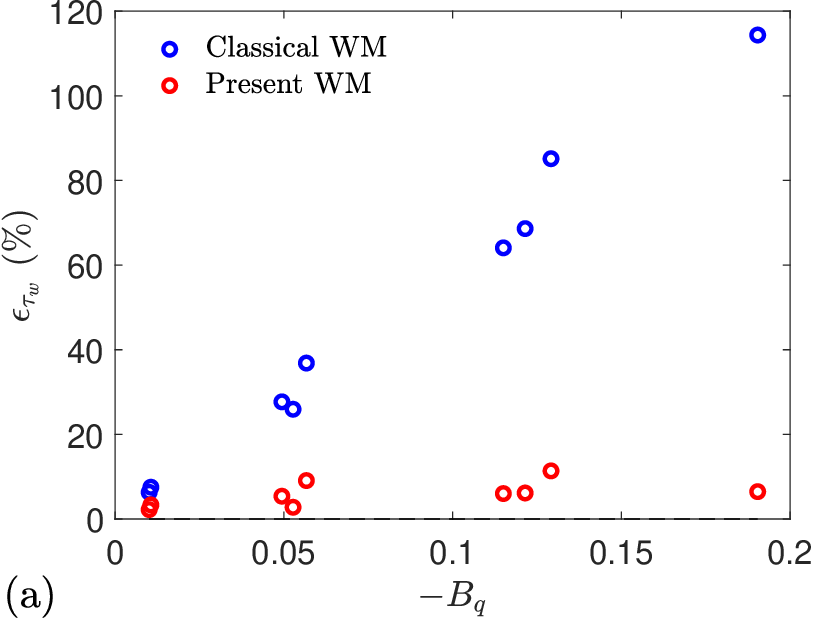}
		\includegraphics[height=0.3\linewidth]{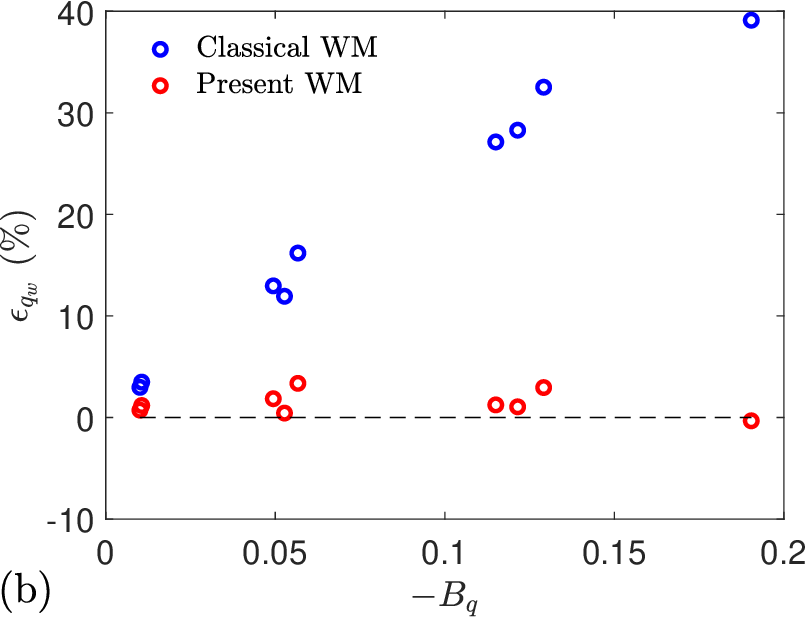}}
	\caption{WMLES {\it a posteriori} modeling errors for the wall shear stress $\tau_w$ (a) and the wall heat flux $q_w$ (b) versus the non-dimensional heat flux $B_q$. WMLES is conducted using the classical (blue) and present (red) wall models for turbulent channel flows at the nine operating conditions listed in table \ref{tab:cases}. 
	}
	\label{fig:les_sweep}
\end{figure}

The velocity and temperature profiles from WMLES are shown in Figure \ref{fig:les_profs_M2} and \ref{fig:les_profs} for turbulent channel flows at four combinations of Reynolds and Mach numbers. In all cases, the present model is significantly more accurate than the classical model for the prediction of velocity and temperature with respect to the reference DNS solutions. For these cases and the others listed in table \ref{tab:cases}, the errors in the predictions of the wall shear stress and the wall heat flux are shown in Figure \ref{fig:les_sweep}. 
The wall model is based on the inversion of the total-stress-based velocity transformation \cite{Griffin2021a} and that was observed to have the greatest improvement over classical approaches in cases with strong heat transfer. This explains why the errors from the classical wall model grow significantly with the strong heat transfer, but the errors from the present model are rather small and do not vary with heat flux.

\begin{figure}
	\centering
	\includegraphics[width=0.4\linewidth]{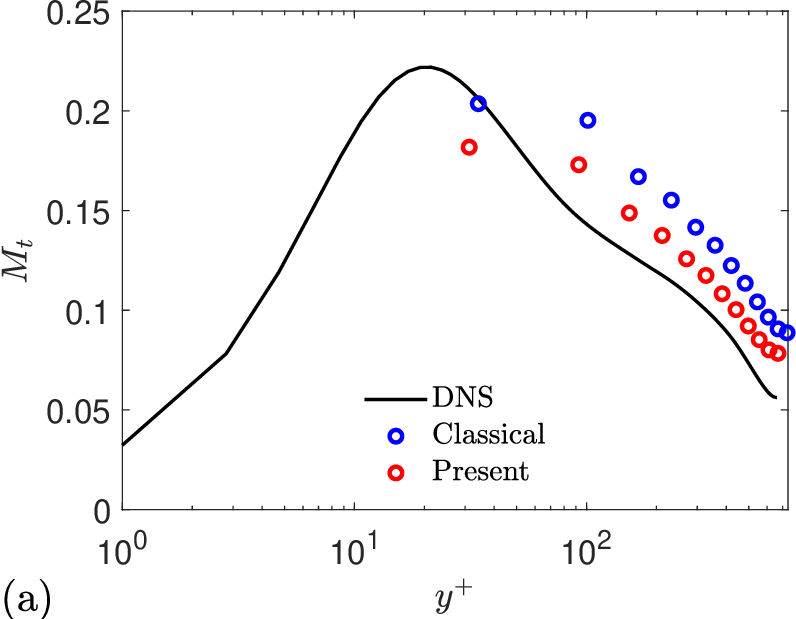}
	\includegraphics[width=0.405\linewidth]{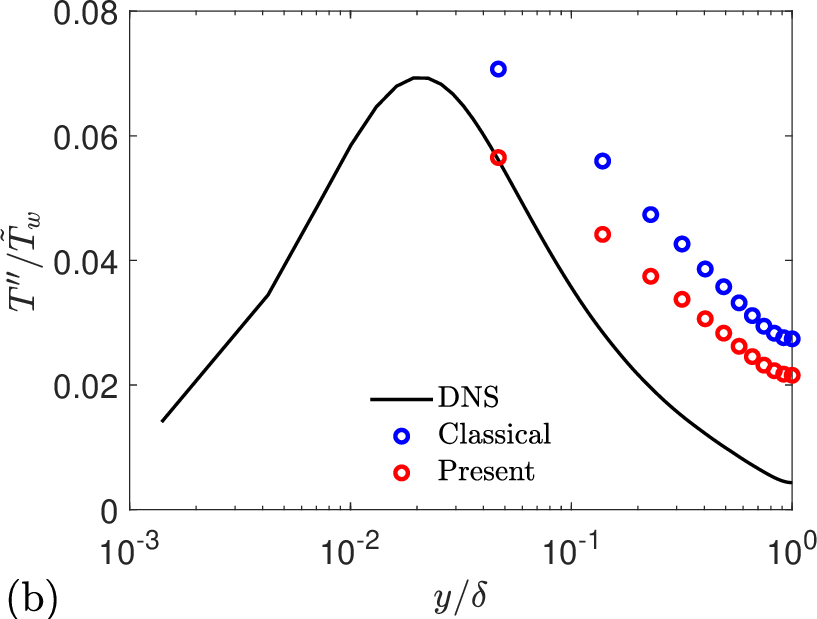}
	\\
	\includegraphics[width=0.4\linewidth]{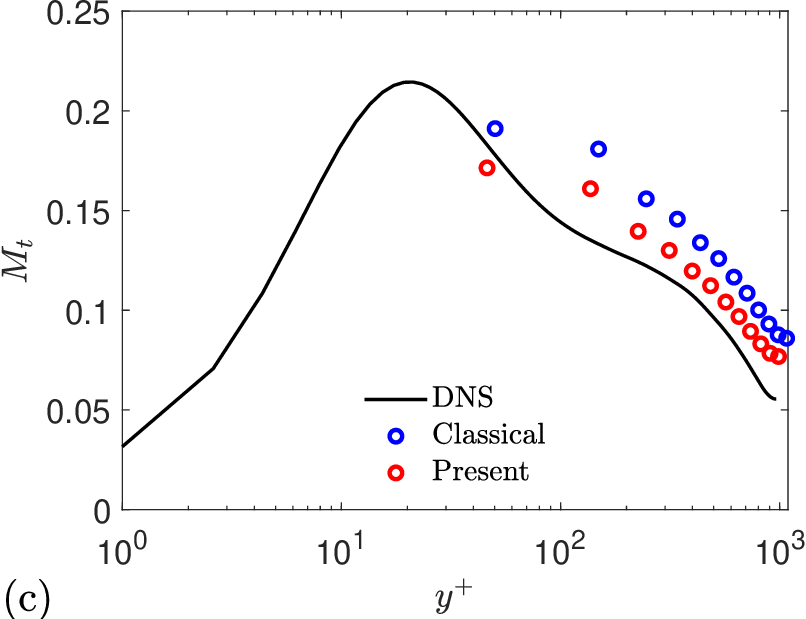}
	\includegraphics[width=0.405\linewidth]{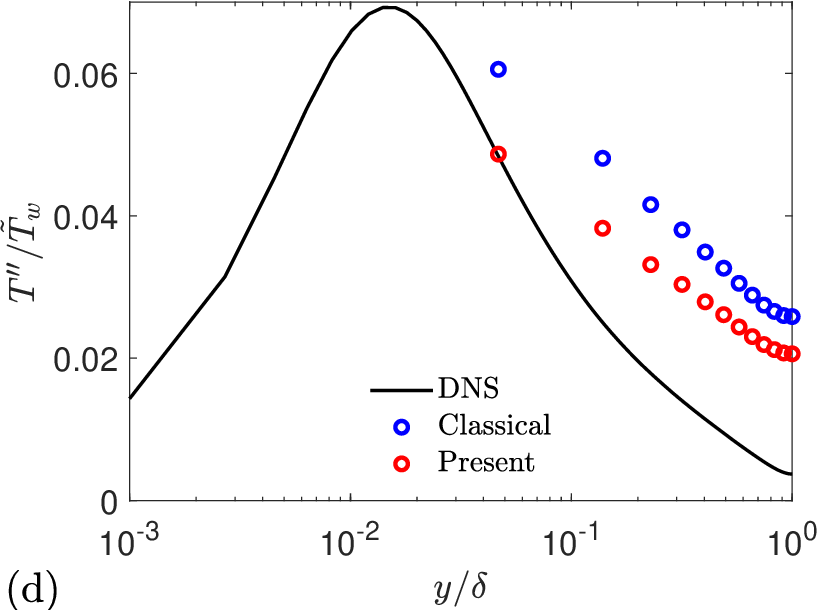}
	\caption{LES turbulent Mach number $M_t$ (a,c) and LES temperature fluctuation $T''$ (b,d) profiles from WMLES with the classical (blue) and present (red) wall models. A channel flow with $M_b=1.7$, $Re_\tau^*=410$, and $-B_q = 0.053$ is shown in panels (a) and (b), and one with $M_b=1.7$, $Re_\tau^*=590$, and $-B_q = 0.049$ is shown in panels (c) and (d). The symbols represent the outer solutions computed by the LES PDE solver.}
	\label{fig:les_profs_fluc_M2}
\end{figure}

\begin{figure}
	\centering
	\includegraphics[width=0.4\linewidth]{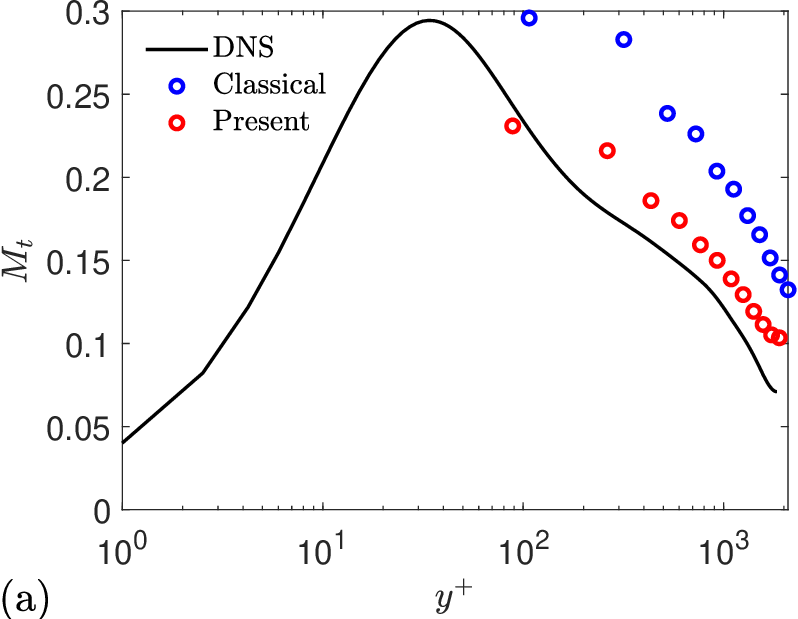}
	\includegraphics[width=0.405\linewidth]{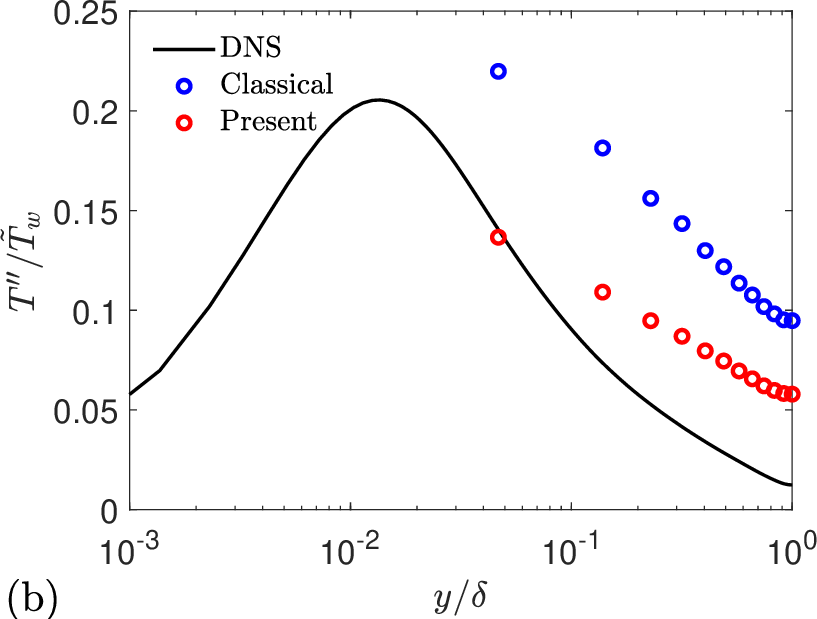}
	\\
	\includegraphics[width=0.4\linewidth]{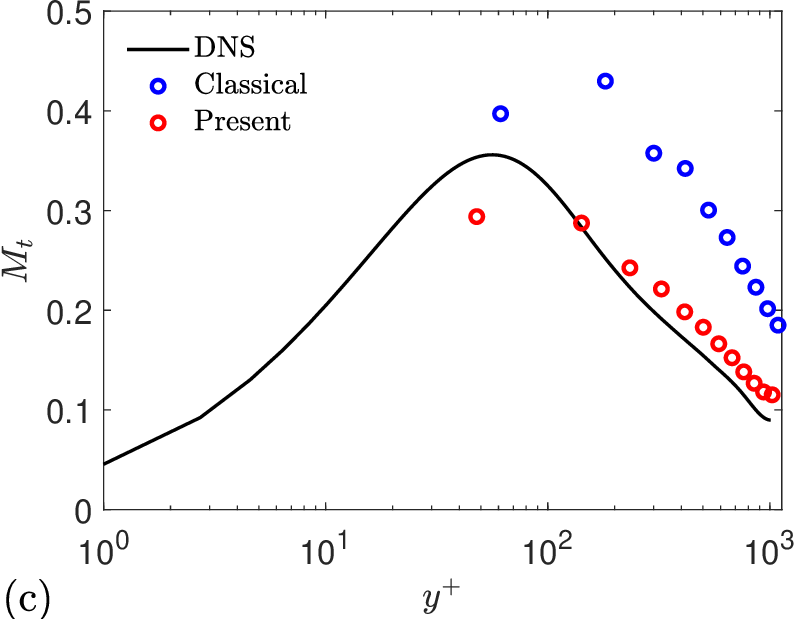}
	\includegraphics[width=0.405\linewidth]{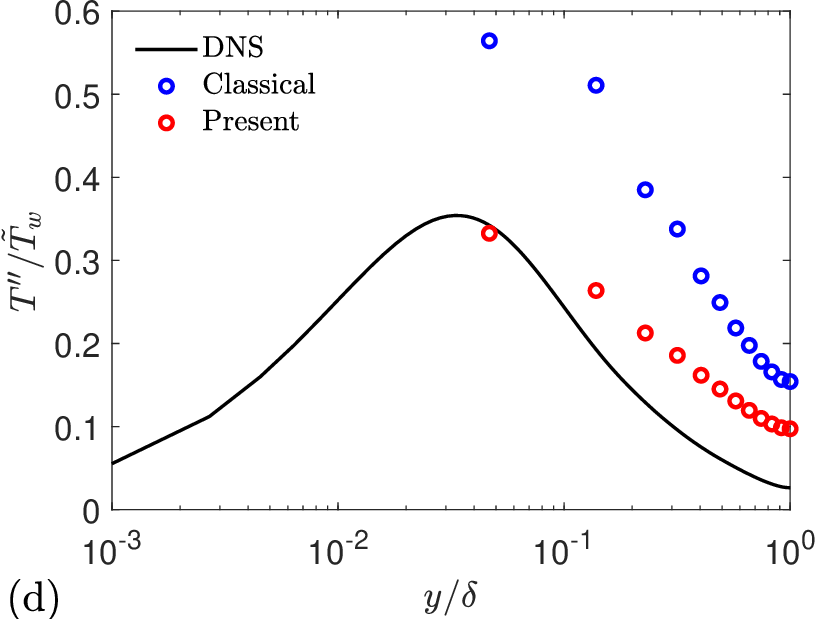}
	\caption{LES turbulent Mach number $M_t$ (a,c) and LES temperature fluctuation $T''$ (b,d) profiles from WMLES with the classical (blue) and present (red) wall models. A channel flow with $M_b=3.0$, $Re_\tau^*=590$, and $-B_q = 0.12$ is shown in panels (a) and (b), and one with $M_b=4.0$, $Re_\tau^*=200$, and $-B_q = 0.19$ is shown in panels (c) and (d). The symbols represent the outer solutions computed by the LES PDE solver.}
	\label{fig:les_profs_fluc}
\end{figure}

The primary quantities of interest for WMLES are the predictions of the mean profiles and fluxes. The fluctuating parts of LES solutions are not expected to exactly agree with DNS results unless the WMLES is conducted with DNS-like resolution, which is impractical. Nevertheless, the effect of wall models on the fluctuating part of the LES solution is presented for comparison between the present and classical models. Figures \ref{fig:les_profs_fluc_M2} and \ref{fig:les_profs_fluc} include profiles of the LES resolved turbulent Mach number $M_t=u''/\sqrt(\gamma R \tilde{T})$ and the LES temperature fluctuations $T''$,  where $(\cdot)''$ denotes the Favre fluctuation $(\cdot)'' = (\cdot) - \tilde{(\cdot)}$. There is an improvement in the predictions of the fluctuating statistics by the present model compared to those by the classical model. An accurate prediction of second-order statistics is unlikely without an accurate prediction of mean statistics. Thus, the improved second-order statistics of the present model are likely a consequence of its improved mean statistics compared to those of the classical model (see Figure \ref{fig:les_profs_M2} and \ref{fig:les_profs}). However, correct prediction of the mean field is not sufficient for the accurate prediction of second-order statistics in LES. In fact, the fluctuations in the LES results are generally over-predicted compared to the DNS data. The over-prediction may be due in part to the wall-blocking effect of stress-based wall model \citep{Bae2018a}. Given the coarse resolution of twelve points across the channel half height, numerical errors and subgrid-scale model errors are certainly contributing. The subgrid-scale model has not been adapted for compressibility other than by accounting for variable properties \citep{Moin1991}. The turbulent Mach numbers are on the order of 0.3, which is sufficiently high that modeling for dilatational dissipation is a promising path to further improvements of the fluctuating statistics in the volume of the LES domain. Such research may be pursued independently of the current study focusing on wall modeling and the prediction of mean profiles and fluxes.

\subsection{Sensitivity to numerical resolution and the matching location}
In WMLES, the wall model exchanges data with the outer LES solver at the matching location. The modeling error in the inner wall modeled equations may grow as the matching distance increases, which motivates placing the matching location near the wall. On the other hand, the matching location should be far enough from the wall in terms of the LES mesh resolution so that the LES solver can resolve the large scales of turbulence at the height of the matching location. Otherwise, numerical errors may contaminate the matching data that is provided as input to the wall model. \cite{Kawai2012} demonstrate this trade-off and how LES numerical errors contaminate the wall-modeled solution if the matching distance is on the order of the wall-normal grid resolution. The optimal matching distance will depend on the accuracy of a specific LES solver, but a typical choice is $y_m \ge 3\Delta$ \citep{Kawai2012}, where $\Delta$ is the wall-normal grid spacing near the wall.

\begin{figure}
	\centerline{
		\includegraphics[width=0.4\linewidth]{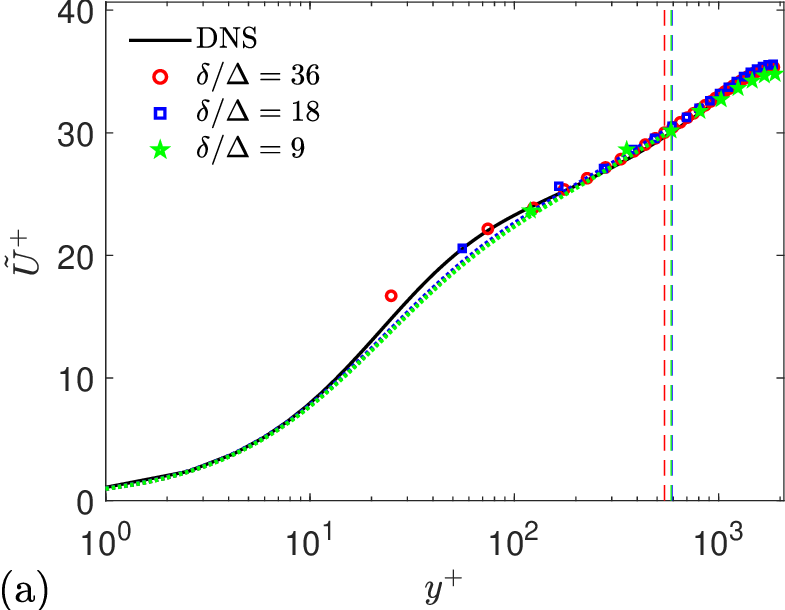}
		\includegraphics[width=0.4\linewidth]{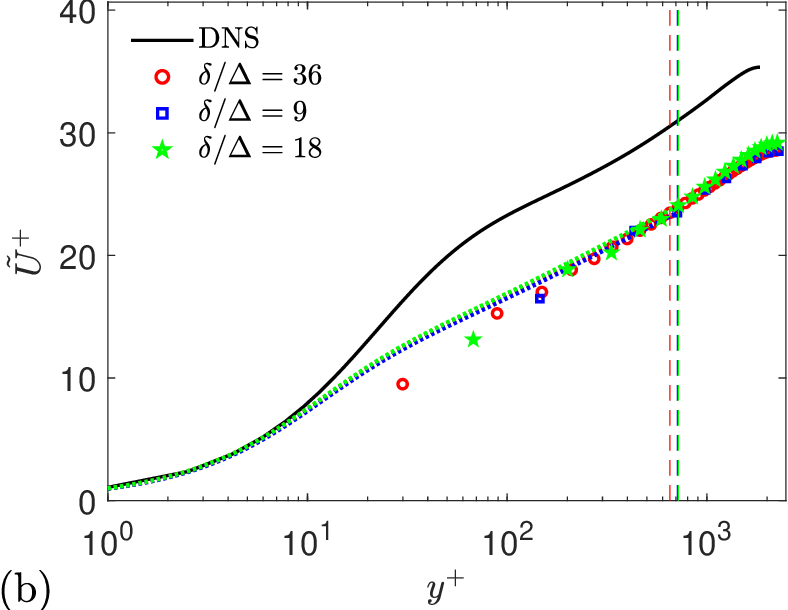}}
	\caption{{\it A posteriori} mesh sensitivity study of a channel flow at $M_b=3.0$ and $Re_\tau^*=590$ with the matching location fixed at $y_m=0.3\delta$ for all cases, as indicated by the vertical dashed lines. The colors indicate the numerical resolution $\Delta$. The outer WMLES solutions and the inner wall-modeled velocity profiles are indicated with symbols and dotted curves, respectively, for the present wall model (a) and the classical wall model (b).}
	\label{fig:conv}
\end{figure}

To evaluate the convergence and sensitivity of the presently proposed wall model, two types of mesh convergence studies are considered. In the first study, the matching location is held fixed at $y_m=0.3\delta$, which corresponds in semi-local units to $y_m^*=186$ and $y_m^*=237$ for the present model and classical model cases across all resolutions.
For the case of $M_b=3.0$ and $Re_\tau=1800$, the numerical resolution of the WMLES is varied. In Figure \ref{fig:conv}, the WMLES solutions are shown for three LES resolutions with 9, 18, and 36 grid points across the channel half-height. The uniform hexagonally close-packed mesh topology with global refinement is employed, resulting in three meshes with $2.0\times10^4$, $1.6\times 10^5$, and $1.3\times 10^6$ control volumes, respectively (note that the reference DNS uses as many as $6.4\times 10^8$ control volumes). 
In this study, the LES numerical errors at the matching location are expected to diminish as the resolution is refined, but modeling errors from using the wall model over the domain $y\in[0,0.3\delta]$ are not expected to change with resolution. For this reason, the classical model shows a large error in the log intercept of the velocity profile that is persistent with refinement and consistent with {\it a priori} analysis in Figure \ref{fig:ap_profs_c2}(a). For the finest resolution with the present model, the grid point nearest to the wall exhibits an error that is persistent with refinement, which is consistent with the observations of \citep{Kawai2012} and does not affect the accuracy of the simulation since the inner solution is applicable for $y<y_m$. For both the present and classical models, the results are only weakly dependent on the grid resolution. This suggests that the leading source of error for the simulations with the classical wall model is in fact the wall model rather than the numerical or subgrid-scale modeling errors, even on the coarsest simulation with 9 grid points per channel half height.

\begin{figure}
	\centerline{
		\includegraphics[width=0.4\linewidth]{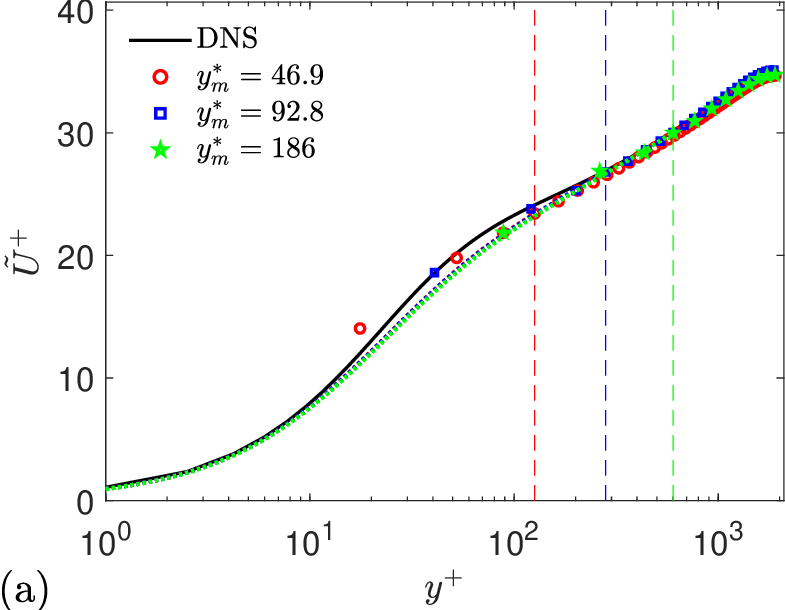}
		\includegraphics[width=0.4\linewidth]{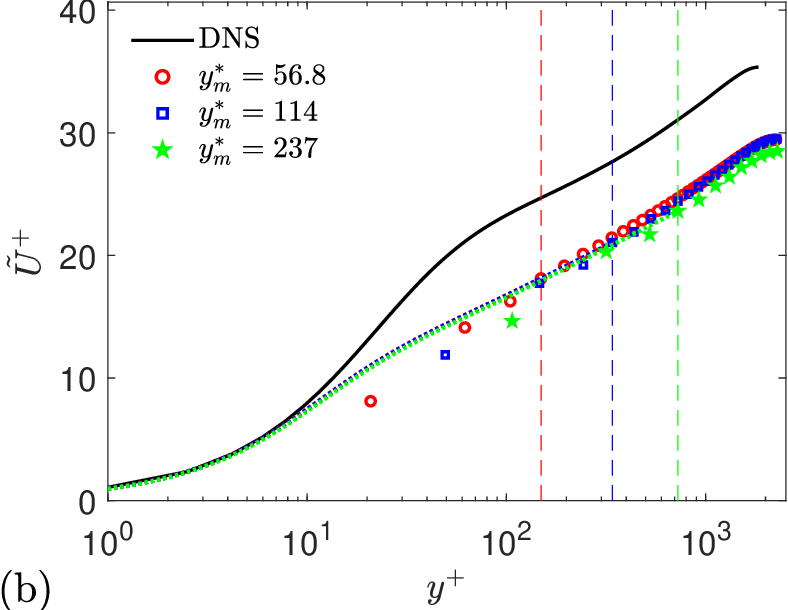}}
	\caption{{\it A posteriori} mesh sensitivity study of a channel flow at $M_b=3.0$ and $Re_\tau^*=590$ with matching locations dependent on the grid resolution as $y_m = 4\Delta$. The colors indicate semi-local matching distance $y_m^*$, which is also indicated with vertical dashed lines. The outer WMLES solutions and the inner wall-modeled velocity profiles are indicated with symbols and dotted curves, respectively, for the present wall model (a) and the classical wall model (b).}
	\label{fig:conv_yvar}
\end{figure}

\begin{figure}
	\centerline{
		\includegraphics[width=0.4\linewidth]{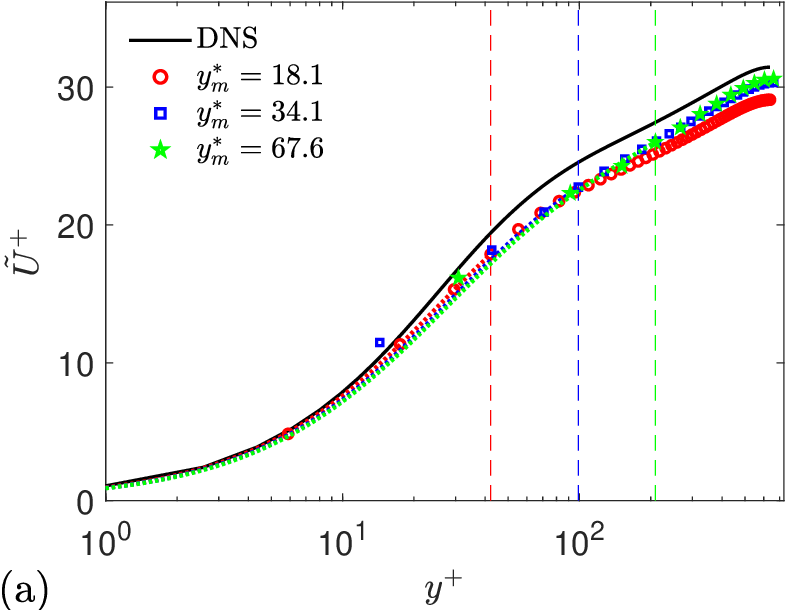}
		\includegraphics[width=0.4\linewidth]{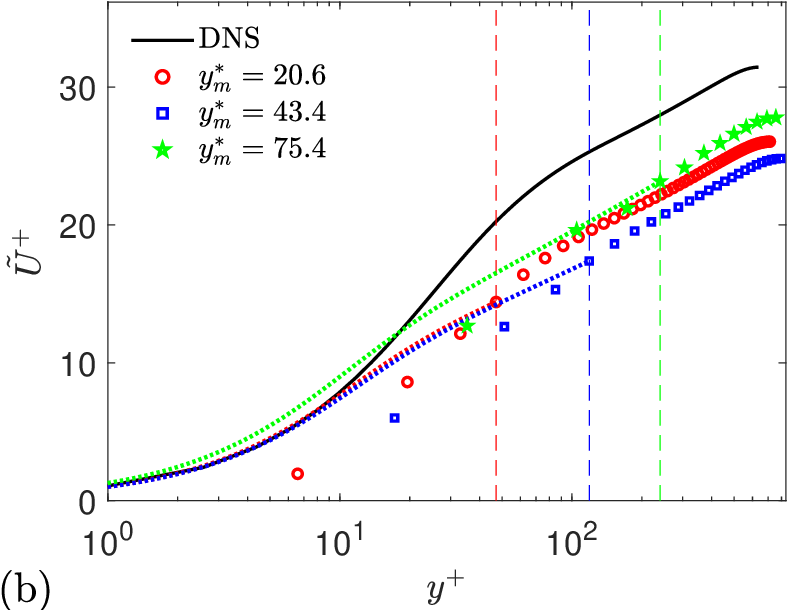}}
	\caption{{\it A posteriori} mesh sensitivity study of a channel flow at $M_b=3.0$ and $Re_\tau^*=200$ with matching locations dependent on the grid resolution as $y_m = 4\Delta$. The colors indicate semi-local matching distance $y_m^*$, which is also indicated with vertical dashed lines. The outer WMLES solutions and the inner wall-modeled velocity profiles are indicated with symbols and dotted curves, respectively, for the present wall model (a) and the classical wall model (b).}
	\label{fig:conv_yvar_R2}
\end{figure}

In the second grid convergence study, the models are tested in the way that WMLES is typically used in practice. That is, the matching distance is moved toward the wall as the grid is refined. In this study, two channel flows with different Reynolds number conditions are considered for three LES resolutions with 12, 24, and 48 grid points across the channel half height. The matching locations are $y_m= 0.3\delta$, $0.15\delta$, and $0.075\delta$, respectively, which corresponds to $y_m = 4 \Delta$ for all cases, thus the effect of near-wall LES numerical errors is expected to be minor \citep{Kawai2012}. In Figure \ref{fig:conv_yvar}, the convergence study is performed for $M_b=3.0$ and $Re_\tau^*=590$, and a lower Reynolds number case of $M_b=3.0$ and $Re_\tau^*=200$ is shown in Figure \ref{fig:conv_yvar_R2}. In both cases, the accuracy of the present model is relatively high and insensitive to mesh resolution compared to that of the classical model. For the higher Reynolds number test, the matching locations in semi-local units are always in the logarithmic region of the boundary layer. Therefore, the WMLES results are not sensitive to refinement over this range of resolutions. However, for the lower Reynolds number case, the most refined meshes lead to semi-local matching locations $y_m^*$ in the buffer region. For the classical model, because the relative error of the modeled $U^+$ versus the DNS $U^+$ is maximal in the region of the buffer layer and early log layer (compare to similar {a priori} results in Figure \ref{fig:ap_profs_ode_coef}), the convergence behavior for the classical model is complex in this regime. In other words, as the mesh is refined, although the LES numerical errors are diminishing, the wall modeling errors for the classical model may increase or decrease depending on the matching location since the relative modeling error does not monotonically reduce with wall-normal distance. On the other hand, the outer solution of the present model is relatively accurate irrespective of the matching location because the inner wall-modeled solution agrees well with the DNS solution throughout the viscous sublayer, buffer layer, and log layer (which is consistent with similar {a priori} results in Figure \ref{fig:ap_profs_ode_coef}).

%
%

\section{Conclusion} \label{sec:conclusion}
In this work, a wall model is proposed for turbulent wall-bounded flows with heat transfer. The model uses an established ODE description of incompressible flow, transforms that equation to account for compressibility effects, and is closed with an algebraic temperature-velocity relation. The resulting model can accurately estimate the near-wall profiles of temperature and velocity when the matching location is in the inner layer. This model is suitable for deployment as a boundary condition for an outer LES or RANS solver, an inflow generation scheme, or the base flow for perturbation methods, possibly with the incompressible model augmented with a wake profile for the outer layer of the boundary layer. The proposed method can only be as accurate as the models on which it is based, namely, the forward velocity transformation and the algebraic temperature-velocity relation. While these models have been widely validated in channel and pipe flows and boundary layers with moderate pressure gradients, further studies in complex flows are warranted, e.g., the developing boundary layers on a blunt body behind a curved shock.

The model is first tested {\it a priori} to verify that it can recover the boundary layer velocity and temperature data when provided with matching data from DNS. Numerical results reveal that the model accurately recovers the targeted profiles well, and the predicted wall stress and heat flux are within a few percent of their expected values for a wide database of DNS data for high-Mach-number turbulent channel flows, pipe flows, and boundary layers (48 cases with edge Mach numbers in the range of 0.77--11 and semi-local friction Reynolds numbers in the range of 170--5700). The model is also tested {\it a posteriori} as a boundary condition for WMLES in turbulent channel flows with bulk Mach numbers $M_b=0.7$--$4.0$ and $Re_\tau=320$--$1800$. Especially in flows with strong heat transfer, the proposed model is substantially more accurate than the classical ODE-based near-wall model. The superior performance of the present model is due to two key differences with respect to the classical model: 1) the constant turbulent Prandtl number assumption is replaced with a more accurate algebraic temperature-velocity relation and 2) the van Driest velocity transformation is replaced with the total-shear-stress velocity transformation \citep{Griffin2021a}.

\section*{Acknowledgments}
Kevin Griffin acknowledges support from the National Defense Science and Engineering Graduate Fellowship, the Stanford Graduate Fellowship, the Stanford Lieberman Fellowship, and the Exascale Computing Project (Grant17-SC-20SC), a collaborative effort of two US Department of Energy organizations (Office of Science and the National Nuclear Security Administration) responsible for the planning and preparation of a capable exascale ecosystem, including software, applications, hardware, advanced system engineering, and early testbed platforms, in support of the nation’s exascale computing imperative. Lin Fu acknowledges funding from the Research Grants Council (RGC) of the Government of Hong Kong Special Administrative Region (HKSAR) with RGC/ECS Project (No. 26200222) and from the Guangdong Basic and Applied Basic Research Foundation (No. 2022A1515011779). Parviz Moin acknowledges support from NASA grant (No. NNX15AU93A).
We wish to gratefully acknowledge helpful comments from Sanjeeb T. Bose.

This work was authored in part by the National Renewable Energy Laboratory, operated by Alliance for Sustainable Energy, LLC, for the U.S. Department of Energy (DOE) under Contract No. DE-AC36-08GO28308. The views expressed in the article do not necessarily represent the views of the DOE or the U.S. Government. The U.S. Government retains and the publisher, by accepting the article for publication, acknowledges that the U.S. Government retains a nonexclusive, paid-up, irrevocable, worldwide license to publish or reproduce the published form of this work, or allow others to do so, for U.S. Government purposes.

\backsection[Declaration of interests]{The authors declare that they do not have any financial or non-financial conflict of interests.}

\backsection[Data availability statement]{The data that support the findings of this study are available from the corresponding authors upon reasonable request. Matlab code implementing the proposed model will be available in the following public repository after the manuscript is accepted for publication:
\href{https://github.com/kevingriffin1/comp_wm}{\url{https://github.com/kevingriffin1/comp_wm}}}

\backsection[Author ORCID]{
Kevin Griffin \href{https://orcid.org/0000-0002-0866-6224}{(0000-0002-0866-6224)};
Lin Fu \href{https://orcid.org/0000-0001-8979-8415}{(0000-0001-8979-8415)}
}

\bibliographystyle{jfm}
\bibliography{comp_database,comp_grid_point,comp_vel_transf,comp_wall,incomp_database,incomp_wall,RANS,1AAA,comp_les}

\end{document}